\documentclass[journal]{IEEEtran}

\usepackage[table,xcdraw]{xcolor}
\usepackage[T1]{fontenc}
\usepackage{subcaption}
\usepackage{color}
\usepackage{varioref}
\usepackage{textcomp}
\usepackage{amsthm}
\usepackage{amsmath}
\usepackage{amssymb}
\usepackage{graphicx}
\usepackage{setspace}
\usepackage{esint}
\usepackage{algorithm}
\usepackage{algpseudocode}
\usepackage{pifont}
\usepackage{wrapfig}
\usepackage{multirow}
\usepackage{tabu}
\usepackage{amsmath}

\usepackage{enumitem}
\usepackage{cite}
\usepackage{cleveref}
\makeatletter

\newcommand{\Rmnum}[1]{\expandafter\@slowromancap\romannumeral #1@}
\makeatother

\newtheorem{theorem}{Theorem}
\newtheorem{lemma}{Lemma}
\newtheorem{remark}{Remark}

\theoremstyle{definition}
\newtheorem{exmp}{Example}

\providecommand{\propositionname}{Proposition}

\ifCLASSINFOpdf

\else

\fi

\usepackage[T1]{fontenc}
\usepackage{etoolbox}

\makeatletter
\patchcmd{\maketitle}{\@fnsymbol}{\@alph}{}{}  % Footnote numbers from symbols to small letters
\makeatother

\title{Computation Scheduling for Distributed Machine Learning with Straggling Workers}
\author{\IEEEauthorblockN{Mohammad Mohammadi Amiri and\thanks{The authors are with the Information Processing and Communications Laboratory (IPC-Lab), Department of Electrical and Electronic Engineering, Imperial College London, London SW7 2AZ, U.K. (e-mail: m.mohammadi-amiri15@imperial.ac.uk; d.gunduz@imperial.ac.uk).}\thanks{This work has been supported by the European Research Council (ERC) through Starting Grant BEACON (agreement No. 677854).}
Deniz G\"und\"uz}

}
\date{}

\begin{document}

\maketitle

%To be considered for the 2017 IEEE Jack Keil Wolf ISIT Student Paper Award. 

%\thispagestyle{empty}
\begin{abstract}
We study scheduling of computation tasks across $n$ workers in a large scale distributed learning problem with the help of a master. Computation and communication delays are assumed to be random, and redundant computations are assigned to workers in order to tolerate stragglers. We consider sequential computation of tasks assigned to a worker, while the result of each computation is sent to the master right after its completion. Each computation round, which can model an iteration of the stochastic gradient descent (SGD) algorithm, is \textit{completed} once the master receives $k$ distinct computations, referred to as the \textit{computation target}. Our goal is to characterize the \textit{average completion time} as a function of the \textit{computation load}, which denotes the portion of the dataset available at each worker, and the computation target. We propose two computation scheduling schemes that specify the tasks assigned to each worker, as well as their computation schedule, i.e., the order of execution. Assuming a general statistical model for computation and communication delays, we derive the average completion time of the proposed schemes. We also establish a lower bound on the minimum average completion time by assuming prior knowledge of the random delays. Experimental results carried out on Amazon EC2 cluster show a significant reduction in the average completion time over existing coded and uncoded computing schemes. It is also shown numerically that the gap between the proposed scheme and the lower bound is relatively small, confirming the efficiency of the proposed scheduling design. 
\end{abstract}

%\begin{IEEEkeywords}
%Gaussian broadcast channels, decentralized caching, superposition coding.
%\end{IEEEkeywords}

%\newpage
\vspace{-.35cm}
\section{Introduction}\label{SecIntro}
The growing computational complexity and memory requirements of emerging machine learning applications involving massive datasets cannot be satisfied on a single machine. Thus, distributed computation across tens or even hundreds of computation servers, called \textit{workers}, has been a topic of great recent interest \cite{DCBoyd,DCChoi}. A major bottleneck in distributed computation is that the overall performance can significantly deteriorate due to slow servers, referred to as \textit{stragglers}. To mitigate the limitation of stragglers, coded computation techniques, inspired by erasure codes against packet losses, have been proposed recently \cite{DCDisMachLearn,DCTandonAlexGD,DCReedSol,DCRaceDutta,DCMomentLDPC,DCAbbeCC}. With coded computation, computations from only a subset of non-straggling workers are sufficient to complete the computation task, thanks to redundant computations performed by the faster workers. In \cite{DCDisMachLearn} the authors employ a maximum-distance separable (MDS) code-inspired distributed computation scheme in a distributed matrix-vector multiplication problem. A more general distributed gradient descent (DGD) problem is considered in \cite{DCTandonAlexGD}, where labeled dataset is distributed across workers, each evaluating the gradient on its own partition. Various coding schemes have been introduced in \cite{DCTandonAlexGD,DCReedSol,DCRaceDutta,DCMomentLDPC,DCAbbeCC}, that assign redundant computations to workers to attain tolerance against stragglers. Coded distributed computation has also been studied for matrix-matrix multiplication, where the labeled data is coded before being delivered to workers \cite{DCYuMaddahMatMult,DCDuttaFahimHadMatMult,QianPolynHighDimMatMat}, and for distributed computing of a polynomial function \cite{DCYuResiliencySecurityPrivacy}. Also, for a linear regression problem, a polynomially coded approach is proposed in \cite{DCLiPolynomialCodedRegression}, where the data is encoded and distributed across the workers to compute the gradient of the loss function. 
%Please see \cite{DCEmrePCMM} for an overview of different approaches.    

Most existing coded computation techniques are designed to tolerate persistent stragglers, and discard computations performed by stragglers. However, in practice we often encounter \textit{non-persistent stragglers}, which, despite being slower, complete a significant portion of the assigned tasks by the time faster workers complete all their tasks \cite{DCHierarchical}. Recently, there have been efforts to exploit the computations carried out by non-persistent stragglers at the expense of increasing the communication load from the workers to the master \cite{DCHierarchical,DCKianiStragglers,DCRateless,DCEmrePCMM,DCLiRandomAssignment}. Techniques studied in \cite{DCHierarchical,DCKianiStragglers,DCRateless,DCEmrePCMM} are based on coding with associated encoding and decoding complexities, which require the availability and processing of all the data points at the \textit{master}. In \cite{DCEmrePCMM} a linear regression problem is studied, and the scheme in \cite{DCLiPolynomialCodedRegression} is extended by allowing each worker to communicate multiple computations sequentially, where the computations are carried out using coded data. The authors in \cite{DCHierarchical} propose to split the computation tasks into multiple levels, and code each level using MDS coding. However, the coding scheme depends on the statistical behavior of the stragglers, which may not be possible to predict accurately in practice. Distributed matrix-vector multiplication is studied in \cite{DCKianiStragglers}. It is shown that, by performing random coding across the dataset, the results can be obtained from a subset of all the tasks assigned to the workers with high probability, where each completes the assigned tasks sequentially. To execute the tasks which are linear functions of their arguments, e.g., matrix-vector multiplication, rateless codes are used in \cite{DCRateless}, requiring a large number of data points assigned to each worker to guarantee decodability of the target function at the master.
%with high probability.

%The tasks are distributed among the workers; each worker sequentially computes and transmits its tasks. 

%This scheme has been extended to a matrix-matrix multiplication task, in which the number of workers is required to be a perfect square number, and the scheduling of the tasks assigned to the workers is designed. 

%We build upon the approach studied in \cite{DCLiRandomAssignment}, which can be regarded as decentralized uncoded distributed computation. 

While significant research efforts have been invested in designing coded computation \cite{DCTandonAlexGD,DCReedSol,DCRaceDutta,DCMomentLDPC,DCAbbeCC,DCYuMaddahMatMult,DCDuttaFahimHadMatMult,QianPolynHighDimMatMat,DCYuResiliencySecurityPrivacy,DCLiPolynomialCodedRegression} techniques, we argue in this paper that uncoded computing and communication can be even more effective in tackling stragglers and reducing the average computation time. We consider computation of an arbitrary function over a dataset, and introduce a centralized scheduling strategy for uncoded distributed computation, where the tasks are assigned to the workers by the master. Each worker can compute a limited number of tasks, referred to as the \textit{computation load}. Computations are carried out sequentially, and the result of each computation is sent to the master right after it is completed. Communication delay from the workers to the master is also taken into account. We assume that both the computation and communication delays are independent across the workers, but may be correlated for different tasks carried out at the same worker. This sequential computation and communication framework allows the master to exploit partial computations by slow workers. The computation is assumed to be completed when the master receives sufficient number of distinct computations, referred to as the \textit{computation target}. Unlike coded computation, uncoded computing approach does not introduce any encoding and decoding delays and complexities; hence, can be particularly efficient for edge learning where the data is inherently distributed \cite{DCMohammadDenizDSGDMACFederated}. It also allows partial decoding, which can be exploited to reduce the communication load for distributed learning \cite{DCKonecnyFederated,DCOneBitQuan,Strom2015ScalableDD}. An uncoded computation approach is also considered in \cite{DCLiRandomAssignment}, where the dataset is split into a limited number of mini-batches, and each worker is randomly assigned a mini-batch of data. This approach requires a large number of workers compared to the number of mini-batches to ensure that the master can recover all the data from the workers with high probability. The authors in \cite{AttiaRaviUncodedDC} study dynamic computation allocation across the workers with feedback providing information about the workers' speeds. The proposed uncoded computation approach in this paper does not impose any constraint on the number of workers, and is designed without any prior knowledge or feedback on the computation and communication delays at the workers.

%We also allow designing the task allocation and scheduling at the workers, and show numerically that it reduces the average computation time significantly.  

The problem under consideration is similar to the well-known job scheduling problem \cite{JobSchedulingGraham}, in which a set of tasks are to be executed by multiple workers given a partial ordering of task execution and the delay associated with each task. The goal is to find a schedule minimizing the total delay, which is shown to be NP-complete \cite{UllmanNPCpmpleteScheduling}. This problem has been studied under different constraints for different applications, such as cloud computing \cite{XueSchedCloudCompACOLB,ZhanPCOCpmpleteScheduling,ReddySchedCloudCompMultiObj}, edge computing \cite{ChoudhariPriorFogComp,YinFogComputeSmartManufacture}, and dispersed computing \cite{PedarsaniSchedulingMultServers,PedarsaniAverstimehrSchedulingComAware}. Our problem differs from the job scheduling one, since no ordering of task execution is imposed, and each task can be executed by an arbitrary number of workers. Also, in our model, the scheduling is designed without having any prior knowledge about the computation and communication delays of the tasks.

Assuming that the computation and communication delays are random variables, our goal is to characterize the minimum \textit{average completion time} as a function of the computation load and computation target. We first provide a generic expression for the average completion time as a function of the \textit{computation schedule}, which specifies both the tasks assigned to each worker and their computation order. We propose two different computation scheduling schemes, and obtain closed-form expressions for their average completion times for a general statistical model of the random delays, which upper bound the minimum average completion time. We also establish a lower bound on the minimum average completion time. The experiments on Amazon EC2 cluster illustrate a substantial reduction in the average completion time with the proposed uncoded computing schemes with task scheduling compared to coded computation schemes and uncoded computation without scheduling of the tasks at the workers \cite{DCLiRandomAssignment}. We highlight that the numerical results are obtained without taking into account the encoding and decoding delays at the master.              
%, which is shown to be tight numerically

%while the communications cost between the workers and the master node is relatively low



The organization of the paper is as follows. We present the system model in Section \ref{SecProbFormul}. In Section \ref{SecDelayAnalysis}, we analyze the performance of the minimum average completion time for the general case. We provide an upper and a lower bound on the minimum average completion time in Section \ref{SecUpperBoundOptAveDelay} and Section \ref{SecLowerBoundOptAveDelay}, respectively. In Section \ref{SecComparison}, we overview some of the alternative approaches in the literature, and compare their performances with the proposed uncoded schemes numerically. Finally, the paper is concluded in Section \ref{SecConc}.

\textit{Notations:} $\mathbb{R}$ and $\mathbb{Z}$ represent sets of real values and integers, respectively. For $i,j \in \mathbb{Z}$, $j \ge i$, $[i:j]$ denotes set $\{ i,i+1, ..., j \}$. For $i \in \mathbb{Z}^+$, we define $[i] \triangleq [ 1:i ]$. $\mathcal{N} \left( 0,\sigma^2 \right)$ denotes a zero-mean normal distribution with variance $\sigma^2$, and, for $a, b \in \mathbb{R}$, $\mathcal{U} \left( a,b \right)$ denotes a uniform distributed over $[a,b]$. $A(i,j)$ represents $(i,j)$-th entry of matrix $A$.
%and finally, $\binom{j}{i}$ returns the binomial  coefficient $j$ choose $i$.
%``$j$ choose $i$''.

\vspace{-.35cm}
\section{System Model}\label{SecProbFormul}
%Each element of the dataset, $X_i$, which we will call as a data point, may correspond to a matrix/vector representing a minibatch of labeled data samples with the same size concatenated along the same dimension.
We consider distributed computation of a function $h$ over a dataset $\boldsymbol{\mathcal{X}}=\left\{ X_1, ..., X_n \right\}$ across $n$ workers. Function $h : \mathbb{V} \to \mathbb{U}$ is an arbitrary function, where $\mathbb{V}$ and $\mathbb{U}$ are two vector spaces over the same field $\mathbb{F}$, and data point $X_i$ is an element of $\mathbb{V}$, $i \in [n]$. The dataset $\boldsymbol{\mathcal{X}}$ is distributed across the workers by the master, and a maximum number of $r \le n$ data points are assigned to each worker, referred to as the \textit{computation load}. We denote by $\mathcal{E}_i$ the indices of the data points assigned to worker $i$, $i \in [n]$, where $\mathcal{E}_i \subset [n]$, $\left| \mathcal{E}_i \right| \le r$. 
%Table \ref{TableNotations} describes the notations and terminologies used in the paper.              

The computations of the tasks assigned to each worker are carried out sequentially. We define the task ordering (TO) matrix $C$ as an $n \times r$ matrix of integers, $C \in [n]^{n \times r}$, specifying the assignment of the tasks to the workers $\boldsymbol{\mathcal{E}} \triangleq \left\{ \mathcal{E}_i \right\}_{i=1}^{n}$, as well as the order these tasks are carried out by each worker $\boldsymbol{\mathcal{O}} \triangleq \left\{ \mathcal{O}_i \right\}_{i=1}^{n}$, where $\mathcal{O}_i$ denotes the computing order of the tasks assigned to worker $i$, $i \in [n]$. Each row of matrix $C$ corresponds to a different worker, and its elements from left to right represent the order of computations. That is, the entry $C(i,j) \in \mathcal{E}_i$ denotes the index of the element of the dataset that is computed by worker $i$ as its $j$-th computation, i.e., worker $i$ first computes $h ( X_{C(i,1)} )$, then computes $h ( X_{C(i,2)} )$, and so on so forth until either it computes $h ( X_{C(i,r)} )$, or it receives the acknowledgement message from the master, and stops computations, $i \in [n]$, $j \in [r]$. Note that the task assignment $\boldsymbol{\mathcal{E}}$ and the order of computations $\boldsymbol{\mathcal{O}}$ are specified by a unique TO matrix $C$. While any $C$ matrix is a valid TO matrix, it is easy to see that the optimal TO matrix will have $r$ distinct entries in each of its rows.

%\footnote{We assume that the processing delay for sending the result of each computation task at each worker is included with the computation delay of that task and is negligible compared to the corresponding computation delay.}
The computations start at time $t=0$ at all the workers, and each worker sends the result of each assigned task to the master right after its computation. We denote the time worker $i$ spends to compute $h \left( X_j \right)$ by $T_{i, j}^{(1)}$, and the communication delay for sending $h \left( X_j \right)$ to the master by $T_{i, j}^{(2)}$, $j \in \mathcal{E}_i$, $i \in [n]$. Thus, the total delay of receiving $h \left( X_j \right)$ from worker $i$ is $T_{i, j}^{(1)} + T_{i, j}^{(2)}$, $j \in \mathcal{E}_i$, $i \in [n]$. If $j \notin \mathcal{E}_i$, we set $T_{i,j}^{(l)} = \infty$, $\forall l \in [2]$, $i \in [n]$. We assume that the computation and communication delays, $T_{i, j}^{(1)}$ and $T_{i, j}^{(2)}$, $\forall i,j \in [n]$, are independent. We further assume that computation (communication) delays at different workers are independent. On the other hand, the computation (communication) delays associated with the tasks at the same worker can be dependent, and we denote the joint cumulative distribution function (CDF) of $T_{i,1}^{(l)}, \dots, T_{i,n}^{(l)}$ by $F_{i, [n]}^{(l)}$, and the joint probability density function (PDF) by $f_{i, [n]}^{(l)}$, $i \in [n]$, $l \in [2]$. We note that the statistical model of the computation (communication) delays at each worker do not depend on any specific order of computing (communicating) tasks, since we assume that the size and complexity of computing (communicating) each data point (computation) is the same.
%The reason for assuming that $T_{i, 1}^{(1)}, \dots, T_{i, n}^{(1)}$ have the same CDF $F_{i}^{(1)}$, for $i \in [n]$, is that the size associated with each data point is assumed to be the same. Thus, the statistics of computation delays for different tasks $h(X_1), \dots, h(X_n)$ are expected to be the same at the same worker. Furthermore, the rationale behind assuming that $T_{i, 1}^{(2)}, \dots, T_{i, n}^{(2)}$ have the same CDF $F_{i}^{(2)}$ is that each worker transmits different computations $h(X_1), \dots, h(X_n)$ with the same data type, e.g., for digital transmission, they are represented with the same number of bits.   

\begin{figure}[t!]
\centering
\includegraphics[width=.9\linewidth]{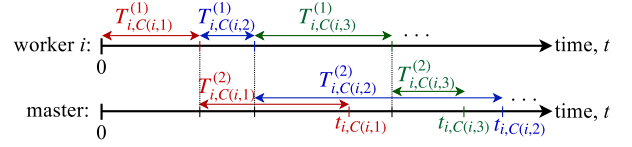}
\caption{Illustration of the computation and communication delays for the computations carried out by worker $i$.}
\label{FigDelays}
\end{figure}

Fig. \ref{FigDelays} illustrates a graphical representation of a realization of the computation and communication delays from worker $i$ to the master. Let $t_{i, j}$ denote the time the master receives $h(X_j)$ from worker $i$, for $i, j \in [n]$, where we set $t_{i, m} = \infty$ if $m \notin \mathcal{E}_i$. Then, the total computation delay of computing $h(X_{C(i,1)})$, $h(X_{C(i,2)}), \dots, h(X_{C(i,j)})$ sequentially plus the communication delay for receiving $h(X_{C(i,j)})$ is
\begin{align}\label{TimetiXjMaster}
t_{i, C(i,j)} = \sum\nolimits_{m=1}^{j} T_{i, C(i,m)}^{(1)} + T_{i, C(i,j)}^{(2)}, \quad i, j \in [n],  
\end{align}
As a result, the master receives computation $h(X_j)$ at time
\begin{align}\label{TimetiMaster}
t_{j} \triangleq \min\nolimits_{i \in [n]} \left\{ t_{i, j} \right\}, \quad j \in [n]   
\end{align}
where the minimization is over the workers.

\begin{exmp}\label{Exmpln3k3r2}
Consider the TO matrix $C$ for $n=4$ and $r=3$:
\begin{align}\label{ExmplMatrixC}
C= \begin{bmatrix}
    1 & 2 & 3 \\
    3 & 2 & 1 \\
    3 & 4 & 1 \\
    4 & 3 & 1
\end{bmatrix},
\end{align}
which dictates the following computation schedule:
\begin{itemize}
\item Worker 1 first computes $h \left( X_1 \right)$, then $h \left( X_2 \right)$, and $h \left( X_3 \right)$.
\item Worker 2 first computes $h \left( X_3 \right)$, then $h \left( X_2 \right)$, and $h \left( X_1 \right)$.
\item Worker 3 first computes $h \left( X_3 \right)$, then $h \left( X_4 \right)$, and $h \left( X_1 \right)$.
\item Worker 4 first computes $h \left( X_4 \right)$, then $h \left( X_3 \right)$, and $h \left( X_1 \right)$.
\end{itemize}
Each worker sends the result of each computation to the master immediately after its completion. Accordingly, we have
\begin{subequations}
\label{Delaysn3k3r2Example}
\begin{align}\label{Delaysn3k3r2ExampleW1}
t_{1, 1} &= T^{(1)}_{1, 1} + T^{(2)}_{1, 1} , \; t_{1, 2} = T^{(1)}_{1, 1} + T^{(1)}_{1, 2} + T^{(2)}_{1, 2}, \nonumber\\ 
t_{1, 3} &= T^{(1)}_{1, 1} + T^{(1)}_{1, 2} + T^{(1)}_{1, 3} + T^{(2)}_{1, 3}, \; t_{1, 4} = \infty, \\
t_{2, 3} &= T^{(1)}_{2, 3} + T^{(2)}_{2, 3} , \; t_{2, 2} = T^{(1)}_{2, 3} + T^{(1)}_{2, 2} + T^{(2)}_{2, 2}, \nonumber\\ 
t_{2, 1} &= T^{(1)}_{2, 3} + T^{(1)}_{2, 2} + T^{(1)}_{2, 1} + T^{(2)}_{2, 1}, \; t_{2, 4} = \infty, 
\label{Delaysn3k3r2ExampleW2}\\
t_{3, 3} &= T^{(1)}_{3, 3} + T^{(2)}_{3, 3} , \; t_{3, 4} = T^{(1)}_{3, 3} + T^{(1)}_{3, 4} + T^{(2)}_{3, 4}, \nonumber\\ 
t_{3, 1} &= T^{(1)}_{3, 3} + T^{(1)}_{3, 4} + T^{(1)}_{3, 1} + T^{(2)}_{3, 1}, \; t_{3, 2} = \infty,
\label{Delaysn3k3r2ExampleW3}\\
t_{4, 4} &= T^{(1)}_{4, 4} + T^{(2)}_{4, 4} , \; t_{4, 3} = T^{(1)}_{4, 4} + T^{(1)}_{4, 3} + T^{(2)}_{4, 3}, \nonumber\\ 
t_{4, 1} &= T^{(1)}_{4, 4} + T^{(1)}_{4, 3} + T^{(1)}_{4, 1} + T^{(2)}_{4, 1}, \; t_{4, 2} = \infty.
\label{Delaysn3k3r2ExampleW4}
\end{align}
%\vspace{-.35cm}
\qed
\end{subequations}
\end{exmp}

For any TO matrix, the computation is considered completed once the master recovers $k$ distinct tasks, referred to as the \textit{computation target}. We allow partial computations, i.e., $k$ can be smaller than $n$. Once the computation target is met, the master sends an acknowledgement message to all the workers to stop computations. Given the TO matrix $C$, we denote the \textit{completion time}; that is, the time it takes the master to receive $k$ distinct computations, by $t_{C}(r,k)$, which is a random variable. We define the average completion time as 
\begin{align}
\overline{t}_{C}(r,k) \triangleq  \mathbb{E} \left[ {t}_{C}(r,k) \right],    
\end{align}
where the randomness is due to the delays. We define the minimum average completion time 
\begin{align}\label{DefAverageDelayComputationCompletion}
\overline{t}^*(r,k) \triangleq \min\nolimits_{C} \left\{ \overline{t}_{C}(r,k) \right\},
\end{align}
where the minimization is taken over all possible TO matrices $C$. The goal is to characterize $\overline{t}^*(r,k)$. 
%It is trivial to see that the optimal TO matrix will have $k$ distinct entries overall.  

\begin{remark}
We have defined each $X_i \in \mathbb{V}$ as a single data point, and assumed that the result of $h(X_i)$ at a worker is transmitted immediately to the master. It is possible to generalize this model by considering $N$ data points instead, with $N \gg n$, and grouping them into $n$ mini-batches, such that each $X_i$ in our model corresponds to a mini-batch of $\left\lceil {N/n} \right\rceil$ data points. A worker sends the average of the gradients for all the data points in a mini-batch after computing all of them. For a mini-batch size of $c$ data points, this corresponds to communicating once every $c$ computations. 
%allowing workers to transmit once every $c > 1$ computations, which can provide a tradeoff between the completion time and the communication load, where $c = r$ corresponds to the approach in coded computing \cite{DCTandonAlexGD,DCReedSol,DCYuMaddahMatMult}. We will instead allow batch computing, where data points are grouped into mini-batches, and are always assigned to a worker together. A worker sends the average of the gradients for all the data points in a mini-batch after computing all of them. For a mini-batch size of $c$ data points, this corresponds to communicating once every $c$ computations. Another approach to extend the proposed model is when there are a total of $N$ data samples, and they are split into $n$ data points $X_1, \dots, X_n$, each consisting of $N/n$ disjoint data samples\footnote{Typically we have $N \gg n$, and we assume that $N$ is divisible by $n$, which can be guaranteed by replicating known data samples.}. After designing the TO matrix $C$ (determining assignment of the data points and the order of their computations at each worker), each worker can communicate with the master once every $c'$ computations with respect to the data samples with the data point assigned to that worker, for $c' \in [N/n]$, where an arbitrary order of computation with respect to the data samples at each data point can be employed. We note that $c' = N/n$ corresponds to the model under consideration.          
\end{remark}

\begin{remark}\label{RemSGDJustif}
Most coded computation schemes in the literature, mainly targeting DGD, require the master to recover the gradients (or, their average) for the whole dataset at each iteration. However, convergence of stochastic gradient descent (SGD) is guaranteed even if the gradient is computed for a random portion of the dataset at each iteration \cite{Strom2015ScalableDD,DCAjiSparse,DCKonecnyFederated,DCSattlerSparseBinary,DCChenAdaComp,DCLinHanDeepGradComp,DCTaoLieSGD,DCWangATOMO}. This is indeed the case for the random straggling model considered here with $k < n$, where the straggling workers; hence, the uncomputed gradients, vary at each iteration.  
\end{remark}

\begin{remark}\label{RemBias}
When $k < n$, in order to prevent bias in the SGD algorithm, we need to make sure that the first $k$ distinct computations received by the master are uniformly random across the mini-batches. If a few workers are significantly faster than the others, we may end up receiving computations corresponding to a few batches assigned to these workers. Alternatively, we can periodically re-index the mini-batches and their corresponding labels randomly after a fixed number of iterations, and provide the workers with the new mini-batches while the TO matrix is fixed. This introduces additional communication from the master to the workers to deliver the missing mini-batches after re-indexing.
\end{remark}

\section{Average Completion Time Analysis}\label{SecDelayAnalysis}

Here we analyze the average completion time $\overline{t}_{C}(r,k)$ for a given TO matrix $C$.

%\begin{prop}\label{CorExactAverageDelayVersusCDF}
%For a TO matrix $C$, let $F_{T_{C}}$ denote the CDF of the random variable $T_{C}(r,k)$. We have
%\begin{align}\label{ExactAverageDelayVersusCDF}
%\overline{T}_{C}(r,k) = \int_0^\infty  {\left( {1 - {F_{T_{C}}}\left( t \right)} \right)dt}.    
%\end{align}
%\end{prop}
%\begin{proof}
%Let $f_{T_{C}}(t) \triangleq d F_{T_{C}}(t)/dt$ be the probability density function (PDF) of ${T}_{C}(r,k)$. Sine $T_{C}(r,k) \ge 0$, i.e., the delay is a non-negative value, we have
%\newcommand\equality{\mathrel{\overset{\makebox[0pt]{\mbox{\normalfont\tiny\sffamily (a)}}}{=}}}
%\begin{align}\label{ProofExactAverageDelayVersusCDF}
%\overline{T}_{C}(r,k) &= \int_0^\infty \tau f_{T_{C}}(\tau) d \tau = \int_0^\infty \int_0^\tau f_{T_{C}}(\tau) d t d \tau \equality \int_0^\infty \int_t^\infty f_{T_{C}}(\tau) d \tau d t \nonumber\\
%&= \int_0^\infty \left( 1 - F_{T_{C}}(t) \right) d t,
%\end{align}
%where (a) follows from changing the order of the integration. 
%\end{proof}
\begin{theorem}\label{TheoremExactAverageDelay}
For a given TO matrix $C$, we have
\begin{align}\label{TheoremExactAverageDelayExpression}
& \Pr \left\{ {t}_{C}(r,k) > t \right\} = 1 - F_{t_{C}}(t) \nonumber\\
& \; \; = \sum\nolimits_{i=n-k+1}^{n} (-1)^{n-k+i+1} \binom{i-1}{n-k} \nonumber\\
& \qquad \qquad \quad \quad \quad \sum\nolimits_{\mathcal{S} \subset [n] : \left| \mathcal{S} \right| = i} \Pr \left\{ t_{j} > t, \forall j \in \mathcal{S} \right\}, 
\end{align}
which yields
\begin{align}\label{RemnGeneralAverageDelay}
& \overline{t}_{C}(r,k) = \sum\nolimits_{i=n-k+1}^{n} (-1)^{n-k+i+1} \binom{i-1}{n-k} \nonumber\\
& \qquad \qquad \; \sum\nolimits_{\mathcal{S} \subset [n] : \left| \mathcal{S} \right| = i} \int_0^\infty \Pr \left\{ t_{j} > t, \forall j \in \mathcal{S} \right\} dt.
\end{align}
\end{theorem}

Note that the dependence of the completion time on the TO matrix in \eqref{TheoremExactAverageDelayExpression} and \eqref{RemnGeneralAverageDelay} is through the statistics of $t_{j}$. % We also note that the expectations in \eqref{TheoremExactAverageDelayExpression} and \eqref{RemnGeneralAverageDelay} are fairly general. 

\begin{proof}
The event $\left\{ {t}_{C}(r,k) > t \right\}$ is equivalent to the union of the events, for which the time to complete any arbitrary set of at least $n-k+1$ distinct computations is greater than $t$, i.e.,
\begin{align}\label{PrTStarGeatertFirstStep}
\Pr \left\{ {t}_{C}(r,k) > t \right\} = \Pr & \bigg\{  \bigcup\nolimits_{\mathcal{G} \subset [n] : n-k+1 \le \left| \mathcal{G} \right| \le n} \Big\{ t_{j} > t,  \nonumber\\
& t_{{j'}} \le t, \forall j \in \mathcal{G} , \forall j' \in \mathcal{G}' \Big\}  \bigg\},    
\end{align}
where we define $\mathcal{G}' \triangleq [n] \backslash \mathcal{G}$. Since the events $\left\{ t_{j} > t, t_{{j'}} \le t, \forall j \in \mathcal{G} , \forall j' \in \mathcal{G}' \right\}$, for all distinct sets ${\cal G} \subset [n]$, are mutually exclusive (pairwise disjoint), we have
\begin{align}\label{PrTStarGeatertSecondStep}
\Pr \left\{ {t}_{C}(r,k) > t \right\} & = \sum\nolimits_{i=n-k+1}^{n} \sum\nolimits_{\mathcal{G} \subset [n]: \left| \mathcal{G} \right| = i} \Pr \Big\{ t_{j} > t,\nonumber\\
& \qquad \qquad \quad t_{{j'}} \le t, \forall j \in \mathcal{G} , \forall j' \in \mathcal{G}' \Big\} \nonumber\\
& = \sum\nolimits_{i=n-k+1}^{n} \sum\nolimits_{\mathcal{G} \subset [n]: \left| \mathcal{G} \right| = i} H_{\mathcal{G},\mathcal{G}'},
\end{align}
where, for $\mathcal{S}_1 \subset [n]$ and $\mathcal{S}_2 \subset [n]$, we define
\begin{align}\label{DefHG}
H_{\mathcal{S}_1,\mathcal{S}_2} \triangleq \Pr \left\{ t_{j_1} > t, t_{{j_2}} \le t, \forall j_1 \in \mathcal{S}_1 , \forall j_2 \in \mathcal{S}_2 \right\}.
\end{align}

\begin{lemma}\label{LemmaHGGprime}
Given a particular set $\mathcal{G} \subset [n]$, $\left| \mathcal{G} \right| = i$, for $i \in [n-k+1:n]$, we have
\begin{align}\label{HGExpressionFori}
& H_{\mathcal{G},\mathcal{G}'} = \sum\nolimits_{m=i}^{n} (-1)^{i+m} \sum\nolimits_{\hat{\mathcal{G}} \subset \mathcal{G}': \left| \hat{\mathcal{G}} \right| = m-i} H_{\mathcal{G} \cup \hat{\mathcal{G}}, \emptyset} \nonumber\\
& = \sum\nolimits_{m=i}^{n} (-1)^{i+m} \sum\nolimits_{\hat{\mathcal{G}} \subset \mathcal{G}': \left| \hat{\mathcal{G}} \right| = m-i} \Pr {\left\{ t_{j} > t, \forall j \in \mathcal{G} \cup \hat{\mathcal{G}} \right\}}.
\end{align}
\end{lemma}
The proof of Lemma \ref{LemmaHGGprime} can be found in Appendix \ref{AppProofInduction}, where we use the fact that, for any $g \in \mathcal{G}'$, we have
\begin{align}\label{HGExpressionForiSimplification}
H_{\mathcal{G},\mathcal{G}'} = H_{\mathcal{G},\mathcal{G}' \backslash \left\{ g \right\} } - H_{ \mathcal{G} \cup \{ g \},\mathcal{G}' \backslash \left\{ g \right\} }.
\end{align}

According to Lemma \ref{LemmaHGGprime}, for $i \in [n-k+1:n]$, we have
\newcommand\equality{\mathrel{\overset{\makebox[0pt]{\mbox{\normalfont\tiny\sffamily (a)}}}{=}}}
\begin{align}\label{HGExpressionForiSumOverG}
& \sum\nolimits_{\mathcal{G} \subset [n]: \left| \mathcal{G} \right| = i} H_{\mathcal{G},\mathcal{G}'} \nonumber\\
& = \sum\nolimits_{\mathcal{G} \subset [n]: \left| \mathcal{G} \right| = i} \sum\nolimits_{m=i}^{n} (-1)^{i+m} \sum\nolimits_{\hat{\mathcal{G}} \subset \mathcal{G}': \left| \hat{\mathcal{G}} \right| = m-i} H_{\mathcal{G} \cup \hat{\mathcal{G}}, \emptyset} \nonumber\\
& = \sum\nolimits_{m=i}^{n} (-1)^{i+m} \sum\nolimits_{\mathcal{G} \subset [n]: \left| \mathcal{G} \right| = i} \sum\nolimits_{\hat{\mathcal{G}} \subset \mathcal{G}': \left| \hat{\mathcal{G}} \right| = m-i} H_{\mathcal{G} \cup \hat{\mathcal{G}}, \emptyset} \nonumber\\
& \equality \sum\nolimits_{m=i}^{n} (-1)^{i+m} \binom{m}{i} \sum\nolimits_{\mathcal{S} \subset [n]: \left| \mathcal{S} \right| = m} H_{\mathcal{S} , \emptyset},
\end{align}
where (a) follows since, for each set $\mathcal{S} = \mathcal{G} \cup \hat{\mathcal{G}}$ with $\left| \mathcal{S} \right| = m$, there are $\binom{m}{i}$ sets $\mathcal{G} \cup \hat{\mathcal{G}}$. Plugging \eqref{HGExpressionForiSumOverG} into \eqref{PrTStarGeatertSecondStep} yields 
\begin{align}\label{PrTStarGeatertPlugging}
& \Pr \left\{ {t}_{C}(r,k) > t \right\} \nonumber\\
& \qquad = \sum\limits_{i=n-k+1}^{n} \sum\limits_{m=i}^{n} (-1)^{i+m} \binom{m}{i} \sum\limits_{\mathcal{S} \subset [n]: \left| \mathcal{S} \right| = m} H_{\mathcal{S} , \emptyset}.
\end{align}
For a particular set $\mathcal{S} \subset [n]$ with $\left| \mathcal{S} \right| = s$, for some $s \in [n-k+1:n]$, the coefficient of $H_{\mathcal{S} , \emptyset}$ in \eqref{PrTStarGeatertPlugging} is given by
\small
\begin{align}\label{CoefficientHS}
&\sum\nolimits_{i=n-k+1}^{s} (-1)^{i+s} \binom{s}{i} \nonumber\\
& = \sum\nolimits_{i=0}^{s} (-1)^{i+s} \binom{s}{i} - \sum\nolimits_{i=0}^{n-k} (-1)^{i+s} \binom{s}{i} \nonumber\\
& = 0 - (-1)^{n-k+s} \binom{s-1}{n-k} = (-1)^{n-k+s+1} \binom{s-1}{n-k},
\end{align}
\normalsize
which results in
\begin{align}\label{PrTStarGeatertLastStep}
&\Pr \left\{ {t}_{C}(r,k) > t \right\}  \nonumber\\
& \quad = \sum\limits_{i=n-k+1}^{n} (-1)^{n-k+i+1} \binom{i-1}{n-k} \sum\limits_{\mathcal{S} \subset [n]: \left| \mathcal{S} \right| = i} H_{\mathcal{S} , \emptyset}.
\end{align}
According to the definition of $H_{\mathcal{S} , \emptyset}$, \eqref{PrTStarGeatertLastStep} concludes the proof of \eqref{TheoremExactAverageDelayExpression}. Furthermore, since $t_{C}(r,k) \ge 0$, we have
\begin{align}\label{ExactAverageDelayVersusCDF}
\overline{t}_{C}(r,k) = \int_0^\infty  {\left( {1 - {F_{t_{C}}}\left( t \right)} \right)dt},    
\end{align}
which yields the expression in \eqref{RemnGeneralAverageDelay}. 
\end{proof}

\begin{remark}\label{RemnEqualk}
For $k=n$, we have
\begin{align}\label{RemnEqualkPr}
&\Pr \left\{ {t}_{C}(r,n) > t \right\} \nonumber\\
& \; = \sum\nolimits_{i=1}^{n} (-1)^{i+1} \sum\nolimits_{\mathcal{S} \subset [n] : \left| \mathcal{S} \right| = i} \Pr \left\{ t_{j} > t, \forall j \in \mathcal{S} \right\}, 
\end{align}
and 
\begin{align}\label{RemnEqualkAverageDelay}
&\overline{t}_{C}(r,n) = \sum\nolimits_{i=1}^{n} (-1)^{i+1} \nonumber\\ 
& \sum\nolimits_{\mathcal{S} \subset [n] : \left| \mathcal{S} \right| = i} \int_0^\infty \Pr \left\{ t_{j} > t, \forall j \in \mathcal{S} \right\} dt.
\end{align}
\end{remark}

%\begin{remark}\label{RemAllAchievSchemesOpt}

The minimum average completion time $\overline{t}^*(r,k)$ can be obtained as a solution of the optimization problem $\overline{t}^*(r,k) = \min_{C} \overline{t}_{C} (r,k)$. Providing a general characterization for $\overline{t}^*(r,k)$ is elusive. In the next section, we will propose two specific computation task assignment and scheduling schemes, and evaluate their average completion times. 

%\end{remark}

\section{Upper Bounds on the Minimum Average Completion Time}\label{SecUpperBoundOptAveDelay}

In this section we introduce two computation task assignment and scheduling schemes, namely \textit{cyclic scheduling} (CS) and \textit{staircase scheduling} (SS). The average completion time for these schemes will provide upper bounds on $\overline{t}^*(r,k)$.

\subsection{Cyclic Scheduling (CS) Scheme}\label{SubSecCS}

The CS scheme is motivated by the symmetry across the workers when we have no prior information on their computation speeds. CS makes sure that each computation task has the same order at different workers. This is achieved by a cyclic shift operator. The TO matrix is given by
\begin{align}\label{MatrixCCSGeneralcCSi}
C_{\rm{CS}} (i,j) = g(i +j-1), \quad \mbox{for $i \in [n]$ and $j \in [r]$},
\end{align} 
where function $g: \mathbb{Z} \to \mathbb{Z}$ is defined as follows:
\begin{align}\label{MatrixCCSFuneDef}
g(m)  \triangleq 
\begin{cases} 
m, &\mbox{if $1 \le m \le n$},\\
m-n, &\mbox{if $m \ge n+1$},\\
m+n, &\mbox{if $m \le 0$}.
\end{cases}
\end{align}
Thus, we have
\begin{align}\label{MatrixCCSGeneral}
C_{\rm{CS}}= \begin{bmatrix}
    g(1) & g(2) & \dots  & g(r) \\
    g(2) & g(3) & \dots  & g(r+1) \\
    \vdots & \vdots & \ddots & \vdots \\
    g(n) & g(n+1) & \dots  & g(n+r-1)
\end{bmatrix},
\end{align}
which, for $i \in [n]$ and $j \in [r]$, results in
\begin{align}\label{TimetiXjMasterCSscheme}
t_{i, g(i+j-1)} = \sum\nolimits_{m=1}^{j} T_{i, g(i+m-1)}^{(1)} + T_{i, g(i+j-1)}^{(2)}.  
\end{align}
For $i \in [n]$, we can re-write \eqref{TimetiXjMasterCSscheme} as follows:
\begin{align}\label{CSSchemeTiXj}
&t_{g(i-j+1),i} \nonumber\\
& = \begin{cases} 
\sum\nolimits_{m=1}^{j} T^{(1)}_{g(i-j+1),g(i-j+m)}+T^{(2)}_{g(i-j+1),i}, & \mbox{if $j \in [r]$},\\
\infty, &\mbox{if $j \notin [r]$},
\end{cases}
\end{align}
which results in
\begin{align}\label{CSSchemeTXj}
t_{i} = \min_{j \in [r]} \left\{ \sum\limits_{m=1}^{j} T^{(1)}_{g(i-j+1),g(i-j+m)}+T^{(2)}_{g(i-j+1),i} \right\}.
\end{align}

\begin{exmp}\label{ExampleCSScheme}
Consider $n=4$ and $r=3$. We have
\begin{align}\label{Exmpln6r4MatrixCCS}
C_{\rm{CS}}= \begin{bmatrix}
    1 & 2 & 3  \\
    2 & 3 & 4  \\
    3 & 4 & 1  \\
    4 & 1 & 2 
\end{bmatrix},
\end{align}
and
\begin{subequations}
\label{DelayXisn6r4Example}
\begin{align}\label{DelayXisn6r4ExampleW1}
t_{1, 1} &= T^{(1)}_{1, 1} + T^{(2)}_{1, 1} , \; t_{1, 2} = T^{(1)}_{1, 1} + T^{(1)}_{1, 2} + T^{(2)}_{1, 2}, \nonumber\\ 
t_{1, 3} &= T^{(1)}_{1, 1} + T^{(1)}_{1, 2} + T^{(1)}_{1, 3} + T^{(2)}_{1, 3}, \; t_{1, 4} = \infty, \\
t_{2, 2} &= T_{2, 2}^{(1)} + T_{2, 2}^{(2)} , \; t_{2, 3} = T_{2, 2}^{(1)} + T_{2, 3}^{(1)} + T_{2, 3}^{(2)}, \nonumber\\ 
t_{2, 4} &= T_{2, 2}^{(1)} + T_{2, 3}^{(1)} + T_{2, 4}^{(1)} + T_{2, 4}^{(2)}, \; t_{2, 1} = \infty, 
\label{DelayXisn6r4ExampleW2}\\
t_{3, 3} &= T_{3, 3}^{(1)} + T_{3, 3}^{(2)} , \; t_{3, 4} = T_{3, 3}^{(1)} + T_{3, 4}^{(1)} + T_{3, 4}^{(2)}, \nonumber\\ 
t_{3, 1} &= T_{3, 3}^{(1)} + T_{3, 4}^{(1)} + T_{3, 1}^{(1)} + T_{3, 1}^{(2)}, \; t_{3, 2} = \infty,
\label{DelayXisn6r4ExampleW3}\\
t_{4, 4} &= T_{4, 4}^{(1)} + T_{4, 4}^{(2)} , \; t_{4, 1} = T_{4, 4}^{(1)} + T_{4, 1}^{(1)} + T_{4, 1}^{(2)}, \nonumber\\ 
t_{4, 2} &= T_{4, 4}^{(1)} + T_{4, 1}^{(1)} + T_{4, 2}^{(1)} + T_{4, 2}^{(2)}, \; t_{4, 3} = \infty,
\label{DelayXisn6r4ExampleW4}
\end{align}\qed
\end{subequations}
\end{exmp}

%Note that we have obtained an explicit characterization of the CS scheme in terms of the CDFs of the computation time of the workers. While this CDF would depend on the particular computation task as well as the capacity and load of the particular server, it is often modeled as a shifted exponential in the literature \cite{DCDisMachLearn,DCHierarchical}.   

%In the following corollary, we characterize $L^{\rm{CS}}_{{\mathcal{S}}}$ for a shifted exponential computation time, i.e., for $i \in [n]$, 
%\begin{align}\label{CSSchemeAveDelayAnalysisCDFFiExp}
%F_i (t) =
%\begin{cases} 
%1 - e^{- \mu_i \left( t- \tau_i \right)}, & \mbox{if $t \ge \tau_i$},\\
%0, & \mbox{if $t < \tau_i$},
%\end{cases}
%\end{align}
%where $\mu_i, \tau_i \in \mathbb{R}^+$. We define $\boldsymbol{\mu} \triangleq \left\{ \mu_1, \dots, \mu_n \right\}$, and $\boldsymbol{\tau} \triangleq \left\{ \tau_1, \dots, \tau_n \right\}$. 

\vspace{-.8cm}
\subsection{Staircase Scheduling (SS) Scheme}\label{SubSecSS}
We can observe that CS imposes the same step size and direction in computations across all the workers. Alternatively, here we propose the SS scheme, which introduces inverse computation orders at the workers. The entries of the TO matrix $C_{\rm{SS}}$ for the SS scheme are given by, for $i \in [n]$, $j \in [r]$,
\begin{align}\label{MatrixCSSGeneralcSSi}
C_{\rm{SS}} (i,j) = g(i+{(-1)^{i-1}} (j-1)).
\end{align} 
It follows that
%\small
\begin{align}\label{MatrixCSSGeneral}
&C_{\rm{SS}}= \nonumber\\
& \begin{bmatrix}
    g(1) & g(2) & \dots  & g(r) \\
    g(2) & g(1) & \dots  & g(3-r) \\
    \vdots & \vdots & \ddots & \vdots \\
    g(n) & g(n+(-1)^{n-1}) & \dots  & g(n+{(-1)^{n-1}}(r-1))
\end{bmatrix},
\end{align}
%\normalsize
which, for $i \in [n]$ and $j \in [r]$, results in
\begin{align}\label{TimetiXjMasterSSscheme}
t_{i, g(i+{(-1)^{i-1}} (j-1))}= & \sum\nolimits_{m=1}^{j} T_{i, g(i+{(-1)^{i-1}} (m-1))}^{(1)}  \nonumber\\
&+ T_{i, g(i+{(-1)^{i-1}} (j-1))}^{(2)}.  
\end{align}
For $i \in [n]$, we can re-write \eqref{TimetiXjMasterSSscheme} as follows:
%\small
\begin{align}\label{SSSchemeTiXj}
&t_{g(i+(-1)^{i+j-1}(j-1)),i} = \nonumber\\
&\begin{cases} 
\sum\limits_{m=1}^{j} T^{(1)}_{g(i+(-1)^{i+j-1}(j-1)),g(i+(-1)^{i+j-1}(j-m))}\\
\qquad \qquad \qquad \quad +T^{(2)}_{g(i+(-1)^{i+j-1}(j-1)),i}, & \mbox{if $j \in [r]$},\\
\infty, &\mbox{if $j \notin [r]$},
\end{cases}
\end{align}
%\normalsize
which results in
\begin{align}\label{SSSchemeTXj}
t_{i} = \min_{j \in [r]} \bigg\{ & \sum\nolimits_{m=1}^{j} T^{(1)}_{g(i+(-1)^{i+j-1}(j-1)),g(i+(-1)^{i+j-1}(j-m))} \Bigg.\nonumber \\
& \Bigg. +T^{(2)}_{g(i+(-1)^{i+j-1}(j-1)),i} \bigg\}.
\end{align}

\begin{exmp}\label{ExampleSSScheme}
Consider $n=4$ and $r=3$. We have
\begin{align}\label{Exmpln6r4MatrixSCS}
C_{\rm{SS}}= \begin{bmatrix}
    1 & 2 & 3  \\
    2 & 1 & 4  \\
    3 & 4 & 1  \\
    4 & 3 & 2 
\end{bmatrix},
\end{align}
and
\begin{subequations}
\label{DelayXisn6r4ExampleSS}
\begin{align}\label{DelayXisn6r4ExampleWSS1}
t_{1, 1} &= T^{(1)}_{1, 1} + T^{(2)}_{1, 1} , \; t_{1, 2} = T^{(1)}_{1, 1} + T^{(1)}_{1, 2} + T^{(2)}_{1, 2}, \nonumber\\ 
t_{1, 3} &= T^{(1)}_{1, 1} + T^{(1)}_{1, 2} + T^{(1)}_{1, 3} + T^{(2)}_{1, 3}, \; t_{1, 4} = \infty, \\
t_{2, 2} &= T_{2, 2}^{(1)} + T_{2, 2}^{(2)} , \; t_{2, 1} = T_{2, 2}^{(1)} + T_{2, 1}^{(1)} + T_{2, 1}^{(2)}, \nonumber\\
t_{2, 4} &= T_{2, 2}^{(1)} + T_{2, 1}^{(1)} + T_{2, 4}^{(1)} + T_{2, 4}^{(2)}, \; t_{2, 3} = \infty, 
\label{DelayXisn6r4ExampleWSS2}\\
t_{3, 3} &= T_{3, 3}^{(1)} + T_{3, 3}^{(2)} , \; t_{3, 4} = T_{3, 3}^{(1)} + T_{3, 4}^{(1)} + T_{3, 4}^{(2)}, \nonumber\\ 
t_{3, 1} &= T_{3, 3}^{(1)} + T_{3, 4}^{(1)} + T_{3, 1}^{(1)} + T_{3, 1}^{(2)}, \; t_{3, 2} = \infty,
\label{DelayXisn6r4ExampleWSS3}\\
t_{4, 4} &= T_{4, 4}^{(1)} + T_{4, 4}^{(2)} , \; t_{4, 3} = T_{4, 4}^{(1)} + T_{4, 3}^{(1)} + T_{4, 3}^{(2)}, \nonumber\\ 
t_{4, 2} &= T_{4, 4}^{(1)} + T_{4, 3}^{(1)} + T_{4, 2}^{(1)} + T_{4, 2}^{(2)}, \; t_{4, 1} = \infty,
\label{DelayXisn6r4ExampleWSS4}
\end{align}
\end{subequations}
\qed
\end{exmp}

\begin{remark}\label{CSSSDifference}
The main difference between CS and SS is that with CS all the workers have the same step size and direction in their computations, while with SS workers with even and odd indices have different directions (ascending and descending, respectively) in the order they carry out the computations assigned to them, but the same step size in their evaluations. 
\end{remark}
%(according to \eqref{MatrixCCSGeneralcCSi})
%(according to \eqref{MatrixCSSGeneralcSSi})   
   
We highlight that the CS and SS schemes may not be the optimal schedules for certain realizations of the straggling behaviour, but our interest is in the average performance. We will see in Section \ref{SecComparison} that both perform reasonably well, and neither scheme outperforms the other at all settings.     
   
%\begin{remark}\label{CSSSAveragePer}
%CS and SS might not provide the optimal scheduling techniques for certain straggling patterns; however, since we study the average completion time, inferiority of a scheme over another one for a specific realization does not guarantee its degradation on average.   
%\end{remark}   

\subsection{Average Completion Time Analysis}\label{SubSecCSSSAnalysis}
Here we analyze the performance of CS and SS providing upper bounds on $\overline{t}^*(r,k)$. We represent the average completion time of CS and SS by $\overline{t}_{\rm{CS}}(r,k)$ and $\overline{t}_{\rm{SS}}(r,k)$, respectively. In order to characterize these average values through \eqref{RemnGeneralAverageDelay}, we need to obtain $H_{\mathcal{S},\emptyset} = \Pr \left\{ t_{i} > t, \forall i \in \mathcal{S} \right\}$, for any set $\mathcal{S} \subset [n]$, $n-k+1 \le \left| \mathcal{S} \right| \le n$, where $t_{1}, \dots, t_{n}$ are given in \eqref{CSSchemeTXj} and \eqref{SSSchemeTXj}, for CS and SS, respectively. For ease of presentation, we denote $H_{\mathcal{S}, \emptyset}$ for CS and SS by $H_{\mathcal{S}, \emptyset}^{{\rm{CS}}}$ and $H_{\mathcal{S}, \emptyset}^{{\rm{SS}}}$, respectively. For $\mathcal{S} \subset [n]$ with $n-k+1 \le \left| \mathcal{S} \right| \le n$, we have 
\begin{align}\label{CSSchemeAveDelayAnalysisHS}
H^{{\rm{CS}}}_{\mathcal{S},\emptyset} &= \Pr \bigg\{ \sum\nolimits_{m=1}^{j} T^{(1)}_{g(i-j+1),g(i-j+m)}+T^{(2)}_{g(i-j+1),i} > t, \bigg. \nonumber\\
& \qquad \qquad \quad \bigg. \forall j \in [r], \forall i \in \mathcal{S} \bigg\} = \Pr \left\{ \mathcal{T}_{\mathcal{S}}^{{\rm{CS}}} (t) \right\},
\end{align}
where we define
\small
\begin{align}\label{CSSchemeAveDelayAnalysisCapTau}
& \mathcal{T}_{\mathcal{S}}^{{\rm{CS}}} (t) \triangleq \Pr \bigg\{ \left( T_{1,1}^{(1)}, \dots, T_{n,n}^{(1)}, T_{1,1}^{(2)}, \dots, T_{n,n}^{(2)} \right): \bigg. \nonumber\\ 
& \bigg. \sum\nolimits_{m=1}^{j} T^{(1)}_{g(i-j+1),g(i-j+m)}+T^{(2)}_{g(i-j+1),i} > t, \forall j \in [r], \forall i \in \mathcal{S} \bigg\}.
\end{align}
\normalsize
Similarly, for any set $\mathcal{S} \subset [n]$, $n-k+1 \le \left| \mathcal{S} \right| \le n$, we have 
\begin{align}\label{SSSchemeAveDelayAnalysisHS}
&H^{{\rm{SS}}}_{\mathcal{S},\emptyset} = \Pr \bigg\{ \sum\nolimits_{m=1}^{j} T^{(1)}_{g(i+(-1)^{i+j-1}(j-1)),g(i+(-1)^{i+j-1}(j-m))} \bigg.\nonumber \\
& \bigg. +T^{(2)}_{g(i+(-1)^{i+j-1}(j-1)),i} > t, \forall j \in [r], \forall i \in \mathcal{S} \bigg\} = \Pr \left\{ \mathcal{T}_{\mathcal{S}}^{{\rm{SS}}} (t) \right\}
\end{align}
where we define
\begin{align}\label{SSSchemeAveDelayAnalysisCapTau}
& \mathcal{T}_{\mathcal{S}}^{{\rm{SS}}} (t) \triangleq \Pr \bigg\{ \left( T_{1,1}^{(1)}, \dots, T_{n,n}^{(1)}, T_{1,1}^{(2)}, \dots, T_{n,n}^{(2)} \right): \bigg. \nonumber\\ 
& \qquad \bigg. \sum\nolimits_{m=1}^{j} T^{(1)}_{g(i+(-1)^{i+j-1}(j-1)),g(i+(-1)^{i+j-1}(j-m))} \nonumber \\
& \qquad +T^{(2)}_{g(i+(-1)^{i+j-1}(j-1)),i} > t, \forall j \in [r], \forall i \in \mathcal{S} \bigg\}.
\end{align}
It follows that, for ${\mathrm{X}} \in \left\{ {\rm{CS}}, {\rm{SS}} \right\}$,
\begin{align}\label{CS_SSchemeAveDelayAnalysisHS}
& H^{{\mathrm{X}}}_{\mathcal{S},\emptyset} = \mathop {\int { \cdots \int } }\limits_{\mathcal{T}_{\mathcal{S}}^{{\mathrm{X}}} (t)} f^{(1)}_{1,[r]} \left( \alpha_1^{(1)} \right) \cdots f^{(1)}_{n,[r]} \left( \alpha_n^{(1)} \right) f^{(2)}_{1,[r]}  \nonumber\\
& \left( \alpha_1^{(2)} \right) \cdots f^{(2)}_{n,[r]} \left( \alpha_n^{(2)} \right)  d \alpha_1^{(1)} \cdots d \alpha_n^{(1)} d \alpha_1^{(2)} \cdots d \alpha_n^{(2)}. 
\end{align}
By plugging \eqref{CS_SSchemeAveDelayAnalysisHS} into \eqref{RemnGeneralAverageDelay}, we can obtain, for ${\mathrm{X}} \in \left\{ {\rm{CS}}, {\rm{SS}} \right\}$,
\begin{align}\label{RemnGeneralAverageDelayCS_SS}
\overline{t}_{{\mathrm{X}}}(r,k) = \sum\nolimits_{i=n-k+1}^{n} & (-1)^{n-k+i+1} \binom{i-1}{n-k} \cdot \nonumber\\
& \sum\nolimits_{\mathcal{S} \subset [n] : \left| \mathcal{S} \right| = i} \int_0^\infty H^{{\mathrm{X}}}_{\mathcal{S},\emptyset} dt.
\end{align}

Note that we have obtained a general characterization of the average completion time of CS and SS in terms of the CDFs of the delays associated with different tasks at different workers. The numerical evaluation of the performances of CS and SS and the lower bound will be presented in Section \ref{SecComparison}.

\section{Lower Bound}\label{SecLowerBoundOptAveDelay}
Here we present a lower bound on $\overline{t}^*(r,k)$ by considering an adaptive model. Note that the TO matrix, in general, may depend on the statistics of the computation and communication delays, i.e., $F_{i,[n]}^{(l)}$, $\forall l \in [2]$, but not on the realization of $T_{i,C(i,j)}^{(l)}$, $i \in [n]$, $j \in [r]$. Let $\hat{T}_{i,j}^{(1)}$ and $\hat{T}_{i,j}^{(2)}$, respectively, represent the computation and communication delays associated with the task worker $i$ executes with its $j$-th computation, $i \in [n]$, $j \in [r]$. We note that $\hat{T}_{i,j}^{(l)}$ is a random variable independent of the TO matrix, $i \in [n]$, $j \in [r]$, $l \in [2]$. We define
\begin{align}\label{LBRealization}
\textbf{T} \triangleq & \left( \hat{T}_{1,1}^{(1)}, \hat{T}_{1,1}^{(2)}, \dots, \hat{T}_{1,r}^{(1)}, \hat{T}_{1,r}^{(2)}, \dots, \hat{T}_{n,1}^{(1)}, \hat{T}_{n,1}^{(2)}, \dots, \right. \nonumber\\
& \;\; \qquad \qquad \qquad \qquad  \qquad \qquad \qquad \; \;\; \left.  \hat{T}_{n,r}^{(1)}, \hat{T}_{n,r}^{(2)} \right).
\end{align}
For each realization of $\textbf{T}$, we allow the master to employ a distinct TO matrix $C_{\textbf{T}}$, and denote the completion time by $t_{C_{\textbf{T}}}(\textbf{T},r,k)$, which is a random variable due to the randomness of $\textbf{T}$. We define 
\begin{align}\label{DefInstantDelayComputationCompletionLB}
{t}_{\rm{LB}}(\textbf{T},r,k) \triangleq \min\nolimits_{C_{\textbf{T}}} \left\{ t_{C_{\textbf{T}}}(\textbf{T},r,k) \right\},
\end{align}
and
\begin{align}\label{DefAverageDelayComputationCompletionLB}
\overline{t}_{\rm{LB}}(r,k) \triangleq  \mathbb{E} \left[ {t}_{\rm{LB}}(\textbf{T},r,k) \right],
\end{align}
where the expectation is taken over $\mathbf{T}$. It is easy to verify that
\begin{align}\label{AverageDelayLowerBounded}
\overline{t}^*(r,k) =& \min\nolimits_{C} \left\{ \mathbb{E} \left[ {t}_{C}(r,k) \right] \right\} \nonumber\\
& \;\;\;\;\;\;\; \ge \mathbb{E} \left[ \min\nolimits_{C_{\textbf{T}}} \left\{ t_{C_{\textbf{T}}}(\textbf{T},r,k) \right\} \right] = \overline{t}_{\rm{LB}}(r,k). 
\end{align}

\begin{remark}\label{RemLBFixedTime}
We remark that ${\textbf{\rm{T}}}$ does not depend on any specific order of computing and communicating tasks; that is, any realization of $\left( \hat{T}_{i,1}^{(1)}, \hat{T}_{i,1}^{(2)}, \dots, \hat{T}_{i,r}^{(1)}, \hat{T}_{i,r}^{(2)} \right)$, i.e., the delays at worker $i$, is independent of any specific value $C(i,j)$ (index of the $j$-th task worker $i$ computes), for $i \in [n], j \in [r]$. This is because we assume that the size and complexity associated with the computation of each data point are the same.    
\end{remark}
%, which means that the computation delay and communication delay of the first task assigned to worker $i$ is $\hat{T}_{i,1}^{(1)}$ and $\hat{T}_{i,1}^{(2)}$, respectively, and so on so forth until the $r$-th task

\begin{figure}[t!]
\centering
\includegraphics[width=1\linewidth]{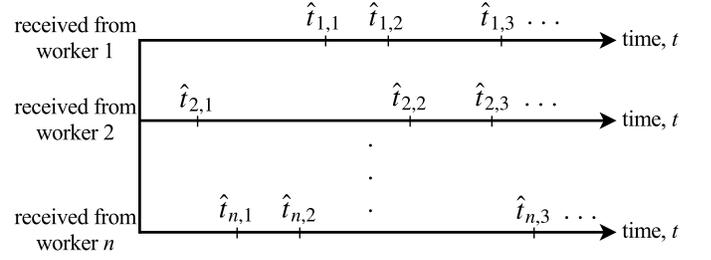}
\caption{Illustration of the arrival times of computations from the workers to the master.}
\label{FigLBDelays}
\end{figure}

We denote by $\hat{t}_{i,j}$ the time at which the master receives the task computed by worker $i$ with its $j$-th computation, $i \in [n]$, $j \in [r]$. It follows that 
\begin{align}\label{TimetiXjMasterLB}
\hat{t}_{i, j} = \sum\nolimits_{l=1}^{j} \hat{T}_{i, l}^{(1)} + \hat{T}_{i, j}^{(2)}.  
\end{align}
Fig. \ref{FigLBDelays} illustrates the time instances computations from the workers are received by the master. For a realization of $\textbf{T}$, ${t}_{\rm{LB}}(\textbf{T},r,k)$ is the $k$-th order statistics of $\left\{ \hat{t}_{1, 1}, \dots, \hat{t}_{1, r}, \dots, \hat{t}_{n, 1}, \dots, \hat{t}_{n, r} \right\}$, i.e., the $k$-th smallest value among $\left\{ \hat{t}_{1, 1}, \dots, \hat{t}_{1, r}, \dots, \hat{t}_{n, 1}, \dots, \hat{t}_{n, r} \right\}$, denoted by $\hat{t}_{\textbf{T}, (k)}$. To prove that ${t}_{\rm{LB}}(\textbf{T},r,k) = \hat{t}_{\textbf{T},{(k)}}$, we note that ${t}_{\rm{LB}}(\textbf{T},r,k)$ cannot be smaller than $\hat{t}_{\textbf{T}, (k)}$, since, according to the definition, for any time before $\hat{t}_{\textbf{T}, (k)}$ master has not received $k$ computations. Also, since master receives the $k$-th computation exactly at time $\hat{t}_{\textbf{T}, (k)}$, knowing the realization of $\textbf{T}$, one can design the TO matrix $C_{\textbf{T}}$ such that the first $k$ computations received by the master are all distinct. Since finding the statistics of $\hat{t}_{\textbf{T}, (k)}$ is analytically elusive, we obtain the lower bound on $\overline{t}^*(r,k)$ through Monte Carlo simulations.

\section{Performance Comparisons}\label{SecComparison}

In this section, we evaluate the average completion time of the proposed CS and SS schemes, and compare them with different results in the literature. We will focus on distributed linear regression as the reference scenario. 

\subsection{Problem Scenario}\label{SubSecProbForm}

We would like to compare the performance of the proposed uncoded computation schemes with coded computation techniques that have received significant interest in recent years. We will consider, in particular, the \textit{polynomially coded} (PC) scheme \cite{DCLiPolynomialCodedRegression} and the \textit{polynomially coded multi-message} (PCMM) scheme \cite{DCEmrePCMM}. PC and PCMM focus exclusively on linear computation tasks; and hence, we also consider a linear regression problem, in which the goal is to minimize 
\begin{align}\label{LinRegLosFuncF}
F \left( \boldsymbol{\theta} \right) = \frac{1}{N} \left\| {\boldsymbol{X} \boldsymbol{\theta}  - \boldsymbol{y}} \right\|_2^2,    
\end{align}
where $\boldsymbol{\theta} \in \mathbb{R}^{d}$ is the model parameter vector, $\boldsymbol{X} \in \mathbb{R}^{N \times d}$ is the data matrix, and $\boldsymbol{y} \in \mathbb{R}^{N}$ is the vector of labels. We split $\boldsymbol{X}$ into $n$ disjoint sub-matrices $\boldsymbol{X} = {\left[ {X_1 \cdots {X_n}} \right]^T}$, where $X_i \in \mathbb{R}^{d \times N/n}$, and $\boldsymbol{y} = {\left[ {y_1^T \cdots {y_n^T}} \right]^T}$, where $y_i \in \mathbb{R}^{N/n}$, $i \in [n]$. The gradient of loss function $F \left( \boldsymbol{\theta} \right)$ is given by
%\small
\begin{align}\label{LinRegGradientLosFuncF}
\nabla F \left( \boldsymbol{\theta} \right) = \frac{2}{N} \boldsymbol{X}^T \left( \boldsymbol{X} \boldsymbol{\theta} - \boldsymbol{y} \right) = \frac{2}{N} \sum\nolimits_{i=1}^{n} \left( X_i X_i^T \boldsymbol{\theta} - X_i y_i \right).      
\end{align}
%\normalsize
We perform gradient descent to minimize \eqref{LinRegLosFuncF}, in which the model parameters at the $l$-th iteration, $\boldsymbol{\theta}_l$, are updated as 
\begin{align}\label{BasicGDModelUpdate}
\boldsymbol{\theta}_{l+1} &= \boldsymbol{\theta}_{l} - \eta_l \cdot \nabla F \left( \boldsymbol{\theta}_l \right) \nonumber\\
&= \boldsymbol{\theta}_{l} - \eta_l \cdot \frac{2}{N} \sum\nolimits_{i=1}^{n} \left( X_i X_i^T \boldsymbol{\theta}_l - X_i y_i \right),   
\end{align}
where $\eta_l$ is the learning rate at iteration $l$. We consider a DGD algorithm, in which the computation of $\nabla F \left( \boldsymbol{\theta} \right)$ is distributed across $n$ workers, and the master updates the parameter vector according to \eqref{BasicGDModelUpdate} after receiving enough computations from the workers, and sends the updated parameter vector to the workers. In the following, we describe the computation tasks carried out by the workers and the master for different schemes, where for the $l$-th iteration, we set 
\begin{align}\label{CSSSRAh}
h (X_i) = X_i X_i^T \boldsymbol{\theta}_l, \quad \mbox{for $i \in [n]$}.     
\end{align}
Since $\boldsymbol{X}^T \boldsymbol{y} = \sum\nolimits_{i=1}^{n} X_i y_i$ remains unchanged over iterations, we assume that its computation is carried out only once by the master node at the beginning of the learning task.

%the loss function

\vspace{-0.307cm}

\subsection{Distributed Computing Schemes}\label{SubSecDisComp}

\underline{\textbf{PC scheme} \cite{DCLiPolynomialCodedRegression}\textbf{:}} At the $l$-th iteration of DGD, the task of computing $\boldsymbol{X}^T \boldsymbol{X} \boldsymbol{\theta}_l = \sum\nolimits_{i=1}^{n} X_i X_i^T \boldsymbol{\theta}_l$ is distributed across the workers. For a computation load $r \ge 2$, worker $i$ stores $r$ distinct matrices ${{\tilde{X}}}_{i,1}, \dots, {{\tilde{X}}}_{i,r}$, where ${{\tilde{X}}}_{i,j} \in \mathbb{R}^{d \times N/n}$ is a linear combination of ${X}_1, \dots, {X}_n$, i.e., ${{\tilde{X}}}_{i,j} = \sum\nolimits_{m=1}^{n} a_{i,j,m} {X}_m$, $a_{i,j,m} \in \mathbb{R}$, $j \in [r]$, $i \in [n]$. Worker $i$, $i \in [n]$, then computes ${{\tilde{X}}}_{i,j} {{\tilde{X}}}_{i,j}^T \boldsymbol{\theta}_l$, $\forall j \in [r]$, and sends their sum, i.e., $\sum\nolimits_{j=1}^{r} {{\tilde{X}}}_{i,j} {{\tilde{X}}}_{i,j}^T \boldsymbol{\theta}_l$, to the master. Thus, having set $h ( {{\tilde{X}}}_{i,j} ) = {{\tilde{X}}}_{i,j} {{\tilde{X}}}_{i,j}^T \boldsymbol{\theta}_l$, the computation task assigned to worker $i$ is $\sum\nolimits_{j=1}^{r} h ( {{\tilde{X}}}_{i,j} )$, and we denote the computation delay at worker $i$ by $T_{{\rm{PC}}, i}^{(1)}$, $i \in [n]$. We also denote the communication delay from worker $i$ to the master by $T_{{\rm{PC}}, i}^{(2)}$, $i \in [n]$. The master receives the computation carried out by worker $i$ at time 
\begin{align}\label{t_PC_i}
t_{{\rm{PC}}, i} = T_{{\rm{PC}}, i}^{(1)} + T_{{\rm{PC}}, i}^{(2)}, \quad \mbox{for $i \in [n]$}. 
\end{align}
The PC scheme allows the master to recover $\boldsymbol{X}^T \boldsymbol{X} \boldsymbol{\theta}_l$ after receiving and processing the results from any $2 \left\lceil {n/r} \right\rceil - 1$ workers \cite{DCLiPolynomialCodedRegression}. Thus, the completion time of PC is the $(2 \left\lceil {n/r} \right\rceil - 1)$-th order statistics of $\{ t_{{\rm{PC}}, 1}, \dots, t_{{\rm{PC}}, n} \}$ denoted by $t_{{\rm{PC}}, (2 \left\lceil {n/r} \right\rceil - 1)}$. The average completion time of the PC is  
\begin{align}\label{AverageDelayPCScheme}
\overline{t}_{\rm{PC}} (r,n) = \mathbb{E} \left[ t_{{\rm{PC}}, (2 \left\lceil {n/r} \right\rceil - 1)} \right],
\end{align}
where the expectation is taken over the computation and communication delay distributions. We note that, for PC, the master needs to further process the received computations to retrieve $\boldsymbol{X}^T \boldsymbol{X} \boldsymbol{\theta}_l$. This additional decoding delay is not taken into account here, but it can be significant.

\begin{exmp}\label{ExamplePCScheme}
Consider $n=4$ and $r=2$. The following matrices are stored at worker $i$, for $i \in [4]$,
\begin{subequations}\label{CopdedDesignPC}
\begin{align}\label{CopdedDesignPC1}
\tilde{X}_{i,1} &= -(i-2)X_1 + (i-1)X_3, \\ 
\tilde{X}_{i,2} &= -(i-2)X_2 + (i-1)X_4.\label{CopdedDesignPC2}
\end{align}
\end{subequations}
Worker $i$, $i \in [4]$, computes $(\tilde{X}_{i,1} \tilde{X}_{i,1}^T + \tilde{X}_{i,2} \tilde{X}_{i,2}^T) \boldsymbol{\theta}_l$, which is equivalent to evaluating a degree-2 polynomial 
\begin{align}\label{PolFunPC}
\phi_1(x) =& (X_1 X_1^T + X_2 X_2^T) \boldsymbol{\theta}_l (x-2)^2 \nonumber\\
& + (X_3 X_3^T + X_4 X_4^T) \boldsymbol{\theta}_l (x-1)^2 \nonumber\\
& - 2 (X_1 X_3^T + X_2 X_4^T) \boldsymbol{\theta}_l (x-1)(x-2)   
\end{align}
at point $x=i$. The master can interpolate polynomial $\phi_1(x)$ after receiving computations from $3$ workers. It then evaluates 
\begin{align}\label{AtMasterPC}
\phi_1(1)+\phi_1(2) = \sum\nolimits_{i=1}^{4} X_i X_i^T \boldsymbol{\theta}_l = \boldsymbol{X}^T \boldsymbol{X} \boldsymbol{\theta}_l.     
\end{align}
\qed
\end{exmp}

\makeatletter
\def\hlinewd#1{%
  \noalign{\ifnum0=`}\fi\hrule \@height #1 \futurelet
   \reserved@a\@xhline}
\makeatother

\underline{\textbf{PCMM scheme} \cite{DCEmrePCMM}\textbf{:}} PC is extended in \cite{DCEmrePCMM} to exploit the partial computations carried out by the workers. For a computation load $r \ge 2$, worker $i$ stores $r$ distinct matrices ${{\hat{X}}}_{i,1}, \dots, {{\hat{X}}}_{i,r}$, where ${{\hat{X}}}_{i,j} = \sum\nolimits_{m=1}^{n} b_{i,j,m} {X}_m$, $b_{i,j,m} \in \mathbb{R}$, $j \in [r]$, $i \in [n]$. Unlike PC, with PCMM proposed in \cite{DCEmrePCMM}, worker $i$, $i \in [n]$, computes $h ( {{\hat{X}}}_{i,1} ), \dots, h ( {{\hat{X}}}_{i,r} )$ sequentially, and sends the result of each computation to the master right after its execution, where $h ( {{\hat{X}}}_{i,j} ) = {{\hat{X}}}_{i,j} {{\hat{X}}}_{i,j}^T \boldsymbol{\theta}_l$. We denote the delay of computing task $h ( {{\hat{X}}}_{i,j} )$ and transmitting the computation to the master by $T_{{\rm{PCMM}}, i, j}^{(1)}$ and $T_{{\rm{PCMM}}, i, j}^{(2)}$, respectively, for $j \in [r]$ and $i \in [n]$. As a result, the master receives computation $h ( {{\hat{X}}}_{i,j} )$, $j \in [r]$, $i \in [n]$, at time
\begin{align}\label{t_PCMM_i}
t_{{\rm{PCMM}}, i, j} = \sum\nolimits_{m=1}^{j} T_{{\rm{PCMM}}, i, m}^{(1)} + T_{{\rm{PCMM}}, i, j}^{(1)}.
\end{align}
It is shown in \cite{DCEmrePCMM} that the master can recover $\boldsymbol{X}^T \boldsymbol{X} \boldsymbol{\theta}_l$ after receiving and processing $2n-1$ computations. Thus, the completion time of PCMM is the $(2 n - 1)$-th order statistics of $\{ t_{{\rm{PCMM}}, i, j}, \forall j \in [r], \forall i \in [n] \}$ denoted by $t_{{\rm{PCMM}}, (2 n - 1)}$. The average completion time of the PC is given by   
\begin{align}\label{AverageDelayPCSchemePCMM}
\overline{t}_{\rm{PCMM}} (r,n) = \mathbb{E} \left[ t_{{\rm{PCMM}}, (2 n - 1)} \right].
\end{align}
Similarly to PC, PCMM also introduces an additional decoding delay, which will be ignored in our numerical comparisons.

\begin{exmp}\label{ExamplePCMMScheme}
Consider $n=4$ and $r=2$. The following matrices are stored at worker $i$, for $i \in [4]$ and $j \in [2]$,
\begin{align}\label{CopdedDesignPCMM}
\hat{X}_{i,j} = \sum\nolimits_{i=1}^{4} X_i \prod\nolimits_{m=1, m \ne i}^{4} \frac{\beta_{i,j}-m}{i-m},
\end{align}
where $\beta_{i,j}$, $\forall i \in [4]$, $\forall j \in [2]$, are different real values. Worker $i$, $i \in [4]$, computes $\hat{X}_{i,1} \hat{X}_{i,1}^T \boldsymbol{\theta}_l$ and $\hat{X}_{i,2} \hat{X}_{i,2}^T \boldsymbol{\theta}_l$ sequentially, and sends the result of each computation right after its completion. Computing $\hat{X}_{i,j} \hat{X}_{i,j}^T \boldsymbol{\theta}_l$, $i \in [4]$, $j \in [2]$, is equivalent to evaluating a degree-6 polynomial 
\begin{align}\label{PolFunPCMM}
\phi_2(x) = & \left( \sum\nolimits_{i=1}^{4} X_i \prod\nolimits_{m=1, m \ne i}^{4} \frac{x-m}{i-m} \right) \nonumber\\
& ~~~ \left( \sum\nolimits_{i=1}^{4} X_i^T \prod\nolimits_{m=1, m \ne i}^{4} \frac{x-m}{i-m} \right) \boldsymbol{\theta}_l  
\end{align}
at point $x=\beta_{i,j}$. The master can interpolate polynomial $\phi_2(x)$ after receiving 7 computations from the workers. Having obtained $\phi_2(x)$, the master evaluates 
\begin{align}\label{AtMasterPCMM}
\sum\nolimits_{i=1}^{4} \phi_2(i) = \sum\nolimits_{i=1}^{4} X_i X_i^T \boldsymbol{\theta}_l = \boldsymbol{X}^T \boldsymbol{X} \boldsymbol{\theta}_l.     
\end{align}
\qed
\end{exmp}

\begin{table*}[t!]
\caption{Characteristics of different schemes under consideration while performing iteration $l$ of DGD.}
\centering
\begin{tabular}{c|c|l|c|c|l}
%\shhline[2pt]
\hlinewd{1.6pt}
Scheme                                    & \begin{tabular}[c]{@{}c@{}}Computation \\ load\end{tabular} & \multicolumn{1}{c|}{Worker $i$}                                                                                                                                                   & \begin{tabular}[c]{@{}c@{}}Computation \\ target\end{tabular} & \begin{tabular}[c]{@{}c@{}}Completion \\ criteria\end{tabular}                                             & \multicolumn{1}{c}{Master}                                                                                                                                                                                                                                                        \\ \hlinewd{1.6pt}
\begin{tabular}[c]{@{}l@{}}$\bullet$ Cyclic \\\quad Scheduling (CS)\\ $\bullet$ Staircase\\ \quad  Scheduling (SS)\\ \\ \; \; ${\rm{X}} \in \{ {\rm{CS}}, {\rm{SS}} \}$\end{tabular} & \multicolumn{1}{l|}{$1 \le r \le n$}                        & \begin{tabular}[c]{@{}l@{}}1. computes \\ \quad $h(X_{C_{\rm{X}}(i,1)})$ \\\quad and sends it\\ 2. computes \\ \quad $h(X_{C_{\rm{X}}(i,2)})$ \\\quad and sends it\\ . . .\end{tabular}           & $1 \le k \le n$                                               & \begin{tabular}[c]{@{}c@{}}receiving\\ $k$ distinct \\ computations\end{tabular}                           & $\boldsymbol{\theta}_{l} - \frac{2 \eta_l}{k} \sum\limits_{i=1}^{k} \left( h \left( X_{p_i} \right) - X_{p_i} y_{p_i} \right)$                                                                                                                                                                                              \\ \hline
\begin{tabular}[c]{@{}c@{}}Random \\ Assignment (RA)\end{tabular}                                        & $r = n$                                                     & \begin{tabular}[c]{@{}l@{}}1. computes \\ \quad $h(X_{C_{\rm{RA}}(i,1)})$ \\\quad and sends it\\ 2. computes \\ \quad $h(X_{C_{\rm{RA}}(i,2)})$ \\\quad and sends it\\ . . .\end{tabular}         & $1 \le k \le n$                                               & \begin{tabular}[c]{@{}c@{}}receiving\\ $k$ distinct \\ computations\end{tabular}                           & $\boldsymbol{\theta}_{l} - \frac{2 \eta_l}{k} \sum\limits_{i=1}^{k} \left( h \left( X_{p_i} \right) - X_{p_i} y_{p_i} \right)$                                                                                                                                                                                                  \\ \hline
\begin{tabular}[c]{@{}c@{}}Polynomially \\ Coded (PC)\end{tabular}                                        & $r \ge 2$                                                   & \begin{tabular}[c]{@{}l@{}}1. computes \\ \quad $\sum\nolimits_{j=1}^{r} h ( {{\tilde{X}}}_{i,j} )$ \\\quad and sends it\end{tabular}                                                   & $k = n$                                                       & \begin{tabular}[c]{@{}c@{}}receiving\\ $2 \left\lceil {n/r} \right\rceil - 1$ \\ computations\end{tabular} & \begin{tabular}[c]{@{}l@{}}1. retrieves $\boldsymbol{X}^T \boldsymbol{X} \boldsymbol{\theta}_l$\\ 2. $\boldsymbol{\theta}_{l} - \frac{2 \eta_l}{N}  \left( \boldsymbol{X}^T \boldsymbol{X} \boldsymbol{\theta}_l - \boldsymbol{X}^T \boldsymbol{y} \right)$\end{tabular} \\ \hline
\begin{tabular}[c]{@{}c@{}}Polynomially \\ Coded \\ Multi-Message\\ (PCMM)\end{tabular}                                      & $r \ge 2$                                                   & \begin{tabular}[c]{@{}l@{}}1. computes \\ \quad $h ( {{\hat{X}}}_{i,1} )$ \\\quad and sends it\\ 2. computes \\ \quad $h ( {{\hat{X}}}_{i,2} )$ \\\quad and sends it\\ . . .\end{tabular} & $k = n$                                                       & \begin{tabular}[c]{@{}c@{}}receiving\\ $2n - 1$ \\ computations\end{tabular}                               & \begin{tabular}[c]{@{}l@{}}1. retrieves $\boldsymbol{X}^T \boldsymbol{X} \boldsymbol{\theta}_l$\\ 2. $\boldsymbol{\theta}_{l} - \frac{2 \eta_l}{N}  \left( \boldsymbol{X}^T \boldsymbol{X} \boldsymbol{\theta}_l - \boldsymbol{X}^T \boldsymbol{y} \right)$\end{tabular} \\ \hlinewd{1.6pt}
\end{tabular}
\label{TableSchemesComparison}
\end{table*}

\underline{\textbf{Uncoded computing:}} Next, we focus on uncoded computation schemes. For a computation target $k$, the master updates the parameter vector after receiving $k$ distinct computations, denoted by $h(X_{p_1}), \dots, h(X_{p_k})$, according to  
\begin{align}\label{BasicGDModelUpdateCSSSk}
\boldsymbol{\theta}_{l+1} = \boldsymbol{\theta}_{l} - \eta_l \cdot \frac{2n}{kN} \sum\nolimits_{i=1}^{k} \left( h \left( X_{p_i} \right) - X_{p_i} y_{p_i} \right),   
\end{align}
where we allow updating the parameter vector with partial computations, and if $k = n$, the update is equivalent to
\begin{align}\label{BasicGDModelUpdateCSSSn}
\boldsymbol{\theta}_{l+1} = \boldsymbol{\theta}_{l} - \eta_l \cdot \frac{2}{N} \sum\nolimits_{i=1}^{n} \left( h \left( X_{i} \right) - X_{i} y_{i} \right).    
\end{align}
If $k < n$, the master stores $X_i y_i$, $\forall i \in [n]$, and at the $l$-th iteration computes $\sum\nolimits_{i=1}^{k} X_{p_i} y_{p_i}$ to update the parameter vector as in \eqref{BasicGDModelUpdateCSSSk}. Whereas, if $k = n$, the master computes $\sum\nolimits_{i=1}^{n} X_{i} y_{i}$ once, and updates the parameter vector as in \eqref{BasicGDModelUpdateCSSSn}. 
%(@Deniz, Is having bias a big issue in training? Do we need to present the following? If yes, shouldn't we present it in Section II as a remark?) Furthermore, when $k < n$, in order to prevent bias, we need to make sure that the first $k$ distinct computations received by the master used to update the parameter vector according to \eqref{BasicGDModelUpdateCSSSk} are uniformly random across the data mini-batches. For this purpose, if a few number of workers are significantly faster than the others, such that the first $k$ distinct computations are mostly sent by them, we can periodically re-index the data mini-batches and their corresponding labels randomly after a fixed number of iterations, and provide the workers with the new mini-batches while the TO matrix is fixed. This introduces additional communication from the master to the workers for transmitting the data points assigned to them by the TO matrix which are missing after re-indexing.             

We first consider the \textit{random assignment} (RA) scheme \cite{DCLiRandomAssignment}. For fairness of comparison, we assume that the training samples are divided into $n$ batches. The computation load $r$ for RA is $r=n$, i.e., the entire dataset is available at each worker. Each worker picks a distinct task (without replacement) independently at random, and sends its computation to the master immediately after its completion. In other words, each row of the TO matrix of RA, denoted by $C_{{\rm{RA}}}$, is a random permutation of vector $[1 \cdots n]$. For a computation target $k$, the master updates the parameter vector according to \eqref{BasicGDModelUpdateCSSSk} after receiving $k$ distinct computations, and the corresponding average completion time is denoted by $\overline{t}_{\rm{RA}} (n,k)$.

\begin{exmp}\label{ExampleRAScheme}
Consider $n=r=4$. Assume the RA scheme results in the following TO matrix: 
\begin{align}\label{Exmpln6r4MatrixCRA}
C_{\rm{RA}}= \begin{bmatrix}
    2 & 1 & 4 & 3  \\
    2 & 4 & 1 & 3  \\
    1 & 4 & 3 & 2  \\
    4 & 3 & 1 & 2  \\
\end{bmatrix}.
\end{align}
We have
\begin{subequations}
\label{DelayXisn6r4ExampleRA}
\begin{align}\label{DelayXisn6r4ExampleWRA1}
t_{1, 2} &= T^{(1)}_{1, 2} + T^{(2)}_{1, 2} , \; t_{1, 1} = T^{(1)}_{1, 2} + T^{(1)}_{1, 1} + T^{(2)}_{1, 1}, \nonumber\\ 
t_{1, 4} &= T^{(1)}_{1, 2} + T^{(1)}_{1, 1} + T^{(1)}_{1, 4} + T^{(2)}_{1, 4}, \nonumber\\
t_{1, 3} &= T^{(1)}_{1, 2} + T^{(1)}_{1, 1} + T^{(1)}_{1, 4} + T^{(1)}_{1, 3} + T^{(2)}_{1, 3}, \\
t_{2, 2} &= T^{(1)}_{2, 2} + T^{(2)}_{2, 2} , \; t_{2, 4} = T^{(1)}_{2, 2} + T^{(1)}_{2, 4} + T^{(2)}_{2, 4}, \nonumber\\ 
t_{2, 1} &= T^{(1)}_{2, 2} + T^{(1)}_{2, 4} + T^{(1)}_{2, 1} + T^{(2)}_{2, 1}, \nonumber\\
t_{2, 3} &= T^{(1)}_{2, 2} + T^{(1)}_{2, 4} + T^{(1)}_{2, 1} + T^{(1)}_{2, 3} + T^{(2)}_{2, 3}, 
\label{DelayXisn6r4ExampleWRA2}\\
t_{3, 1} &= T^{(1)}_{3, 1} + T^{(2)}_{3, 1} , \; t_{3, 4} = T^{(1)}_{3, 1} + T^{(1)}_{3, 4} + T^{(2)}_{3, 4}, \nonumber\\ 
t_{3, 3} &= T^{(1)}_{3, 1} + T^{(1)}_{3, 4} + T^{(1)}_{3, 3} + T^{(2)}_{3, 3}, \nonumber\\
t_{3, 2} &= T^{(1)}_{3, 1} + T^{(1)}_{3, 4} + T^{(1)}_{3, 3} + T^{(1)}_{3, 2} + T^{(2)}_{3, 2},
\label{DelayXisn6r4ExampleWRA3}\\
t_{4, 4} &= T^{(1)}_{4, 4} + T^{(2)}_{4, 4} , \; t_{4, 3} = T^{(1)}_{4, 4} + T^{(1)}_{4, 3} + T^{(2)}_{4, 3}, \nonumber\\
t_{4, 1} &= T^{(1)}_{4, 4} + T^{(1)}_{4, 3} + T^{(1)}_{4, 1} + T^{(2)}_{4, 1}, \nonumber\\
t_{4, 2} &= T^{(1)}_{4, 4} + T^{(1)}_{4, 3} + T^{(1)}_{4, 1} + T^{(1)}_{4, 2} + T^{(2)}_{4, 2},
\label{DelayXisn6r4ExampleWRA4}
\end{align}\qed
\end{subequations}
\end{exmp}

\begin{figure*}[t!]
\centering
\begin{subfigure}{.5\textwidth}
  \centering
  \includegraphics[scale=0.55,trim={8pt 6pt 40pt 35pt},clip]{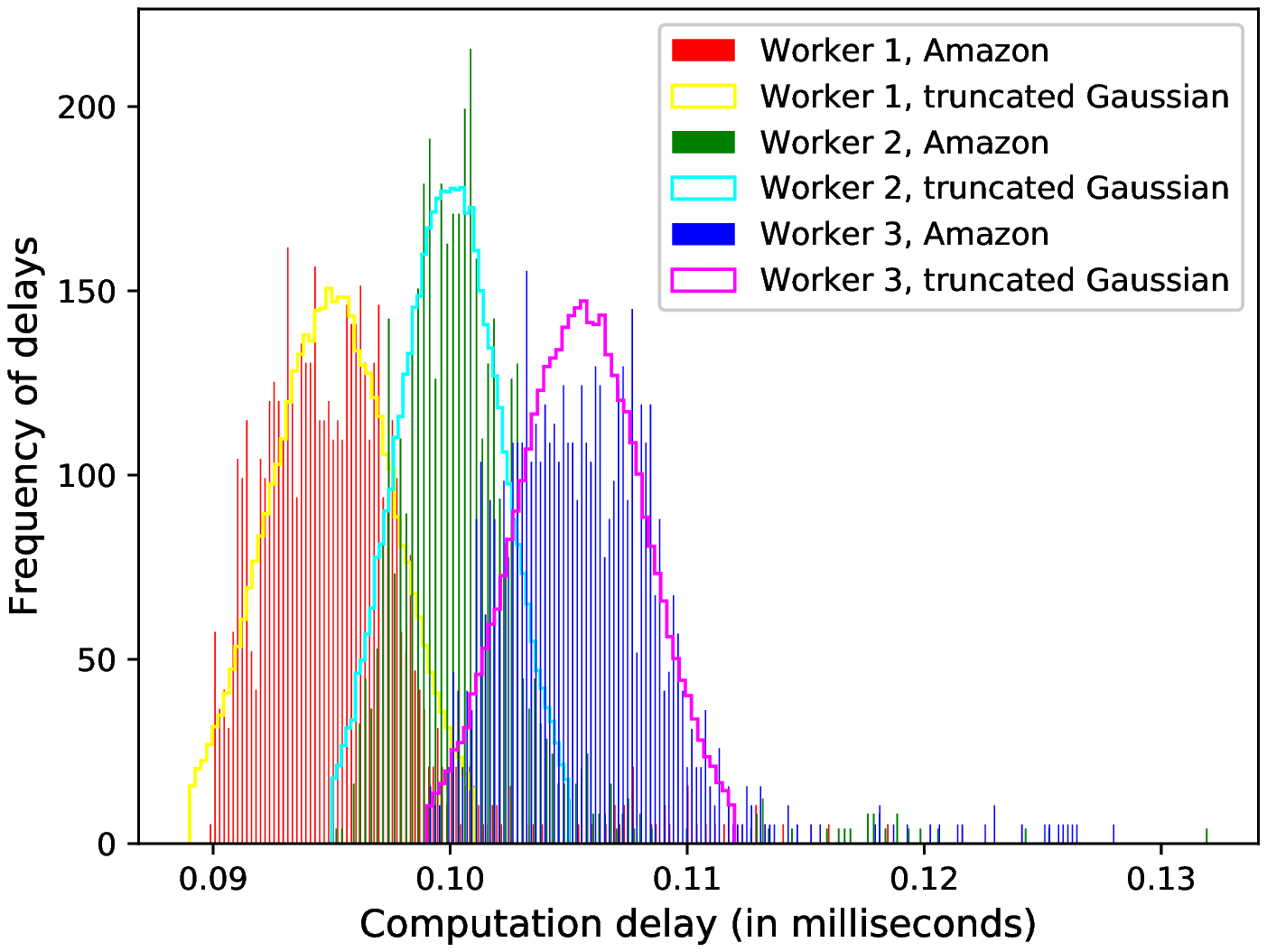}
  %\caption{$\bar{P} = \bar{P}_1$}
  \label{Hist_Compt}
\end{subfigure}%
\begin{subfigure}{.5\textwidth}
  \centering
  \includegraphics[scale=0.55,trim={8pt 6pt 40pt 35pt},clip]{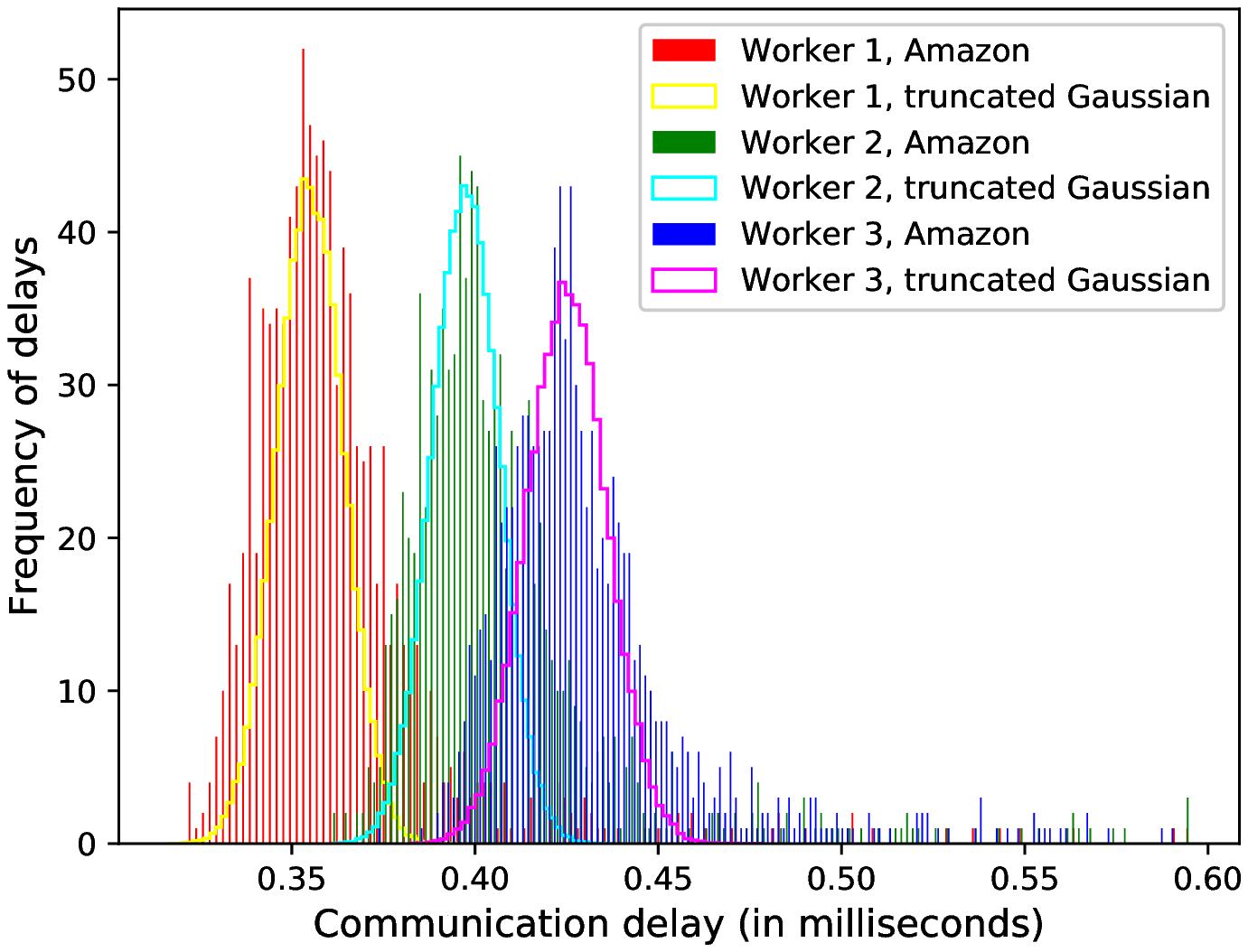}
  %\caption{$\bar{P} = \bar{P}_2$}
  \label{Hist_Comm}
\end{subfigure}
\caption{Histogram of computation and communication delays of three different workers.}
\label{Fig_Histogram}
\end{figure*}

The TO matrices for CS and SS are given in \eqref{MatrixCCSGeneral} and \eqref{MatrixCSSGeneral}, respectively. At the $l$-th iteration of the DGD, the parameter vector is updated by the master according to \eqref{BasicGDModelUpdateCSSSk}.

\begin{remark}\label{RemMaxVecMult}
An alternative approach to tackle the linear regression problem is that, besides $\boldsymbol{X}^T \boldsymbol{y}$, the master computes $\boldsymbol{W} \triangleq \boldsymbol{X}^T \boldsymbol{X}$ once at the beginning of the learning task. Accordingly, the problem reduces to computing matrix-vector multiplication $\boldsymbol{W} \boldsymbol{\theta}_l$ at the $l$-th iteration in a distributed manner, and after recovering $\boldsymbol{W} \boldsymbol{\theta}_l$ the master updates the model parameter vector by 
\begin{align}\label{BasicGDModelUpdateNewApp}
\boldsymbol{\theta}_{l+1} &= \boldsymbol{\theta}_{l} - \eta_l \cdot \frac{2}{N}  \left( \boldsymbol{W} \boldsymbol{\theta}_l - \boldsymbol{X}^T \boldsymbol{y} \right).   
\end{align}
In this case, the proposed CS and SS schemes can be updated accordingly to compute $\boldsymbol{W} \boldsymbol{\theta}_l$ in a distributed manner. To the best of our knowledge, the coded computing scheme tolerating the highest number of straggling workers for the problem of computing $\boldsymbol{W} \boldsymbol{\theta}_l$ distributively is the one proposed in \cite{QianPolynHighDimMatMat}. Due to limited space we do not present the results for this setting here; however, the proposed CS and SS schemes outperform the coded computing scheme in \cite{QianPolynHighDimMatMat} and approach the lower bound similarly to the results presented next.  
%However, this approach requires a significant precoding overhead due to evaluating matrix $\boldsymbol{W}$ as the dimensions of $\boldsymbol{X}$ are typically very high.     
\end{remark}

\begin{figure*}[t!]
\centering
\begin{subfigure}{.5\textwidth}
  \centering
  \includegraphics[scale=0.55,trim={8pt 6pt 40pt 35pt},clip]{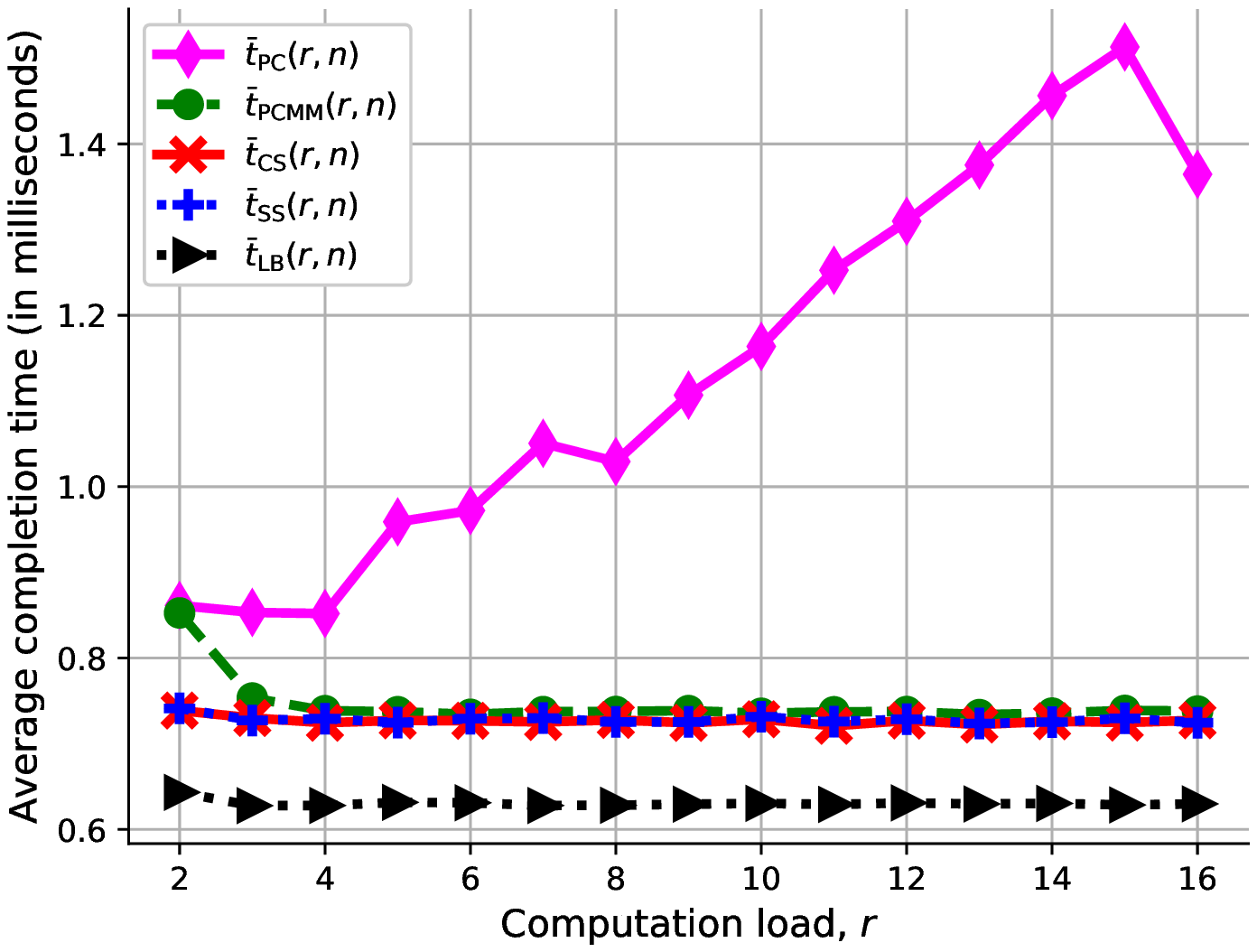}
  \caption{Scenario 1}
  \label{Fig_Exp_Dist_Scenario1}
\end{subfigure}%
\begin{subfigure}{.5\textwidth}
  \centering
  \includegraphics[scale=0.55,trim={8pt 6pt 40pt 35pt},clip]{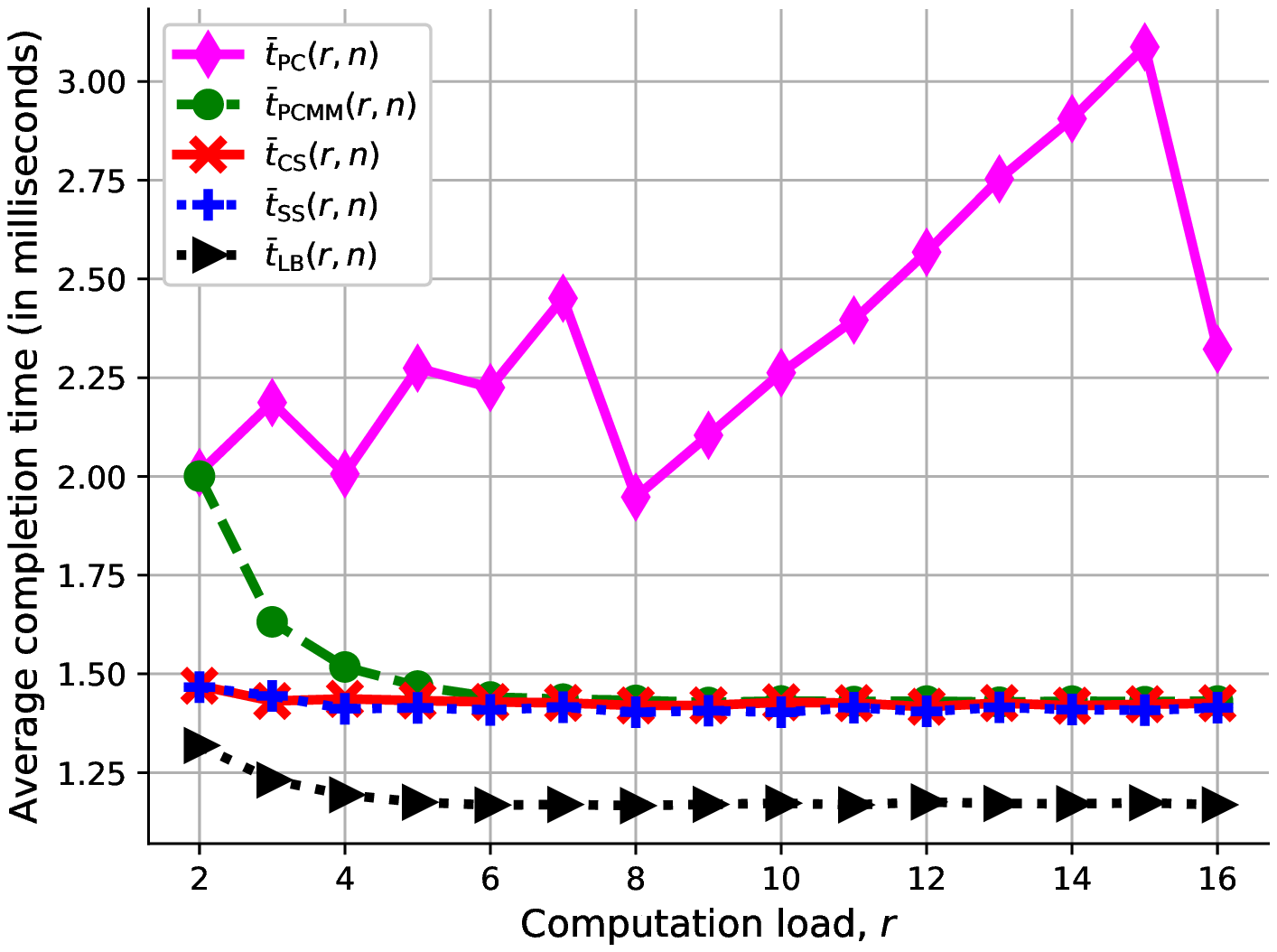}
  \caption{Scenario 2}
  \label{Fig_Exp_Dist_Scenario2}
\end{subfigure}
\caption{Average completion time versus computation load, $r \ge 2$, for the truncated Gaussian delay model in \eqref{ShiftExpModel}.}
\label{Fig_Exp_Dist_Scenario12}
\end{figure*}

We have summarized the characteristics of each of the schemes considered above in Table \ref{TableSchemesComparison}, including the computation tasks conducted by the workers and the master. We note that the computational complexity of the CS, SS, PCMM, and RA schemes at the worker is the same, since with each of the schemes workers need to perform matrix-matrix-vector multiplications sequentially with dimensions $d \times N/n$, $N/n \times d$, and $d$, respectively. In the PC scheme, in addition to the matrix-matrix-vector multiplications, each worker needs to sum its results, $r$ vectors of dimension $d$. With all the schemes, the master first needs to retrieve $\boldsymbol{X}^T \boldsymbol{X} \boldsymbol{\theta}_l$, and then computes $\boldsymbol{\theta}_{l} - \frac{2 \eta_l}{N}  ( \boldsymbol{X}^T \boldsymbol{X} \boldsymbol{\theta}_l - \boldsymbol{X}^T \boldsymbol{y} )$. With the CS, SS, and RA schemes, the master retrieves $\boldsymbol{X}^T \boldsymbol{X} \boldsymbol{\theta}_l$ by adding $n$ $d$-dimensional vectors $h(X_1), \dots, h(X_n)$ (for fairness, we assume $k = n$), which can be done in an online fashion as the computations from the workers arrive. With the PC scheme, the master should wait until it receives $2 \left\lceil {n/r} \right\rceil - 1$ computations, and, in order to retrieve $\boldsymbol{X}^T \boldsymbol{X} \boldsymbol{\theta}_l$, it needs to interpolate $d$ polynomials of degree $2 \left\lceil {n/r} \right\rceil - 2$, then evaluate it at $\left\lceil {n/r} \right\rceil$ points, and finally sum up the results from these $\left\lceil {n/r} \right\rceil$ points, each a vector of dimension $N/n$. Also, with PCMM, the master should first receive $2 n -1$ computations, then interpolate $d$ polynomials of degree $2 n - 2$, next evaluate it at $n$ points, and finally sum up the results from these $n$ points, each a vector of dimension $N/n$, to retrieve $\boldsymbol{X}^T \boldsymbol{X} \boldsymbol{\theta}_l$. Accordingly, it can be concluded that the computational complexity at the master for the coded computing schemes is much higher than the uncoded ones. It is worth noting that, for this study, we do not take into account the computation delay at the master while evaluating the average completion time.

\subsection{Numerical Experiments}\label{SubSecParSet}
For the numerical experiments we generate each entry of data matrix $\boldsymbol{X}$ independently according to distribution $\mathcal{N} \left( 0, 1 \right)$. We also generate the labels as $y_i = (X_i + Z)^T U$, where $Z \in \mathbb{R}^{d \times N/n}$, with each entry distributed independently according to $\mathcal{N} \left( 0, 0.01 \right)$, and $U \in \mathbb{R}^{d}$ with each entry distributed independently according to $\mathcal{U} (0,1)$. For fairness we use the same dataset for all the schemes.

We train a linear regression model using the DGD algorithm described above with a constant learning rate $\eta_l = 0.01$. We run experiments on an Amazon EC2 cluster over t2.micro instance with $n+1$ servers, where one of the servers is designated as the master and the rest serve as workers. We implement different schemes in Python and employ MPI4py library for message passing between different nodes.

At each iteration of the DGD algorithm, we measure the computation and communication delays of each task at each worker. We can then obtain the completion time of each scheme according to its completion criteria. We obtain the average completion time over $500$ iterations. 
%For the lower bound, we find realizations of delays $\hat{T}^{(m)}_{i,j}$, $\forall i \in [n]$, $\forall j \in [r]$, $\forall m \in [2]$, from the delays found by performing the SS scheme. Thus, the lower bound on the average time is associated with the performance of the SS scheme.     

%We denote by $\overline{T}_{\rm{PCMM}} (r,n)$, the average completion time of the PCMM scheme. We note that the PCMM scheme studies an extreme case of sending the computation results sequentially compared to the PC scheme, in which each worker sends only a single message, the sum of all its computations.  

%\subsection{Numerical Experiments}\label{SubSecNumRes}
In Fig. \ref{Fig_Histogram}, we investigate the histograms of computation and communication delays experienced by three different workers. We carried out the experiment on an Amazon EC2 cluster with $N=900$, $d=500$, $n=3$, and set $r=1$ and $k=n$ so that the master waits until it receives the computations from all the workers. Observe that both the computation and communication delays are not highly skewed across different workers. We also plotted the quantized PDF of a truncated Gaussian distribution modelling the delays at each worker. As it can be seen, truncated Gaussian distribution provides a reasonably good estimate of the statistics of both computation and communication delays at different workers. Note that the communication delay at each worker is on average much higher than its computation delay, which verifies that the communication is the major bottleneck in distributed computation and learning \cite{DCMohammadDenizDSGDMACFederated,Strom2015ScalableDD,DCAjiSparse,DCKonecnyFederated,DCSattlerSparseBinary,DCChenAdaComp,DCLinHanDeepGradComp,DCTaoLieSGD,DCWangATOMO}.

We first evaluate and compare the performances of different schemes assuming that both the computation and communication delays follow truncated Gaussian distributions. For simplicity, we assume that the computation and communication delays of different tasks at the same worker are independent, i.e., $f_{i, [n]}^{(l)} = \prod\nolimits_{j=1}^{n} f^{(l)}_{i,j}$, where $f^{(l)}_{i,j}$ denotes the PDF of $T^{(l)}_{i,j}$, for $i,j \in [n]$, $l \in [2]$. For $t \in [\mu_i^{(l)}-a_i^{(l)}, \mu_i^{(l)}+b_i^{(l)}]$, $f^{(l)}_{i,j}$ is given by
\begin{subequations}\label{ShiftExpModel}
\begin{align}
f_{i,j}^{(l)} (t) = \frac{\phi \left( (t-\mu_i^{(l)})/\sigma_i^{(l)} \right)}{\sigma_i^{(l)} \left( \Phi \left( b_i^{(l)}/\sigma_i^{(l)} \right) - \Phi \left( -a_i^{(l)}/\sigma_i^{(l)} \right) \right)},
\end{align}
and $f_{i,j}^{(l)} (t)=0$, otherwise, for $i, j \in [n]$, $l \in [2]$, where
\begin{align}
\phi (t) & \triangleq \frac{1}{\sqrt{2 \pi}} e^{-t^2 / 2},\\
\Phi (t) & \triangleq \frac{1}{2} \left( 1+{\rm{erf}}(t/\sqrt{2}) \right).
\end{align}
\end{subequations}
For PC, we assume that $T^{(1)}_{{\rm{PC}}, i}$ follows the same PDF as $\sum\nolimits_{j=1}^{r} T^{(1)}_{i,j}$, for $i \in [n]$, where there is no loss of generality since the delays are independent and identically distributed (i.i.d.). We further assume that delays $T^{(1)}_{{\rm{PCMM}}, i, j'}$, $\forall j' \in [r]$, follow PDF $f_{i,j}^{(1)}$, $j \in [n]$, for $i \in [n]$. Also, $T^{(2)}_{{\rm{PC}}, i}$ and $T^{(2)}_{{\rm{PCMM}}, i, j'}$, $\forall j' \in [r]$, have PDF $f_{i,j}^{(2)}$, $j \in [n]$, for $i \in [n]$.  
%$\mu_i^{(l)}$ is the decay rate modelling the straggling behaviour of the worker, and $\tau_i^{(l)}$ stands for the processing delay. 

For simplicity, we assume symmetric distributions for the delays, where $a_i^{(l)} = b_i^{(l)}$, $i \in [n], l \in [2]$. We consider two scenarios in our simulations, and for both scenarios we set $a_i^{(1)}=3E5$, $\sigma_i^{(1)}=1E4$, $a_i^{(2)}=2E4$, and $\sigma_i^{(2)}=2E4$, $\forall i \in [n]$, where, for $\alpha, \beta \in \mathbb{R}$, we used the notation $\alpha E \beta$ to denote $\alpha \times 10^{-\beta}$. In Scenario 1, we set $\mu^{(1)}_i = 1E4$ and $\mu^{(2)}_i = 5E4$, $\forall i \in [n]$. In Scenario 2, $\{ \mu_1^{(1)}, \dots, \mu_n^{(1)} \}$ is set as a random permutation of set $\{ 1E4, \frac{4}{3}E4, \dots, \frac{2+n}{3}E4 \}$, and $\{ \mu_1^{(2)}, \dots, \mu_n^{(2)} \}$ is a random permutation of set $\{ 5E4, 5.5E4, \dots, \frac{9+n}{2}E4 \}$. We note that, compared to Scenario 1, the computation and communication delays across the workers are more diverse in Scenario 2. In Fig. \ref{Fig_Exp_Dist_Scenario12} we compare the performances of different schemes for the truncated Gaussian model with $n=16$ workers and $k=n$. For both scenarios, SS slightly improves upon CS, and both CS and SS schemes outperform the coded schemes PC and PCMM for the whole range of $r$. PCMM performs better than PC, and the improvement is less pronounced in Scenario 2, in which the delays are more diverse. Also, compared to Scenario 1, the gap between CS/SS and the coded schemes is less in Scenario 2, and for small $r$ values, the superiority of CS/SS over PCMM is more pronounced. For $r=n$, $\bar{t}_{{\rm{RA}}} (n,n) = 0.86$ and $\bar{t}_{{\rm{RA}}} (n,n) = 1.64$ milliseconds in Scenarios 1 and 2, respectively. SS reduces these average delays by $\% 19.45$ and $\% 16.32$ for Scenarios 1 and 2, respectively, showing that an efficient computation schedule for uncoded computing can reduce the latency.

\begin{figure}[t!]
\centering
\includegraphics[scale=0.6,trim={8pt 6pt 45pt 35pt},clip]{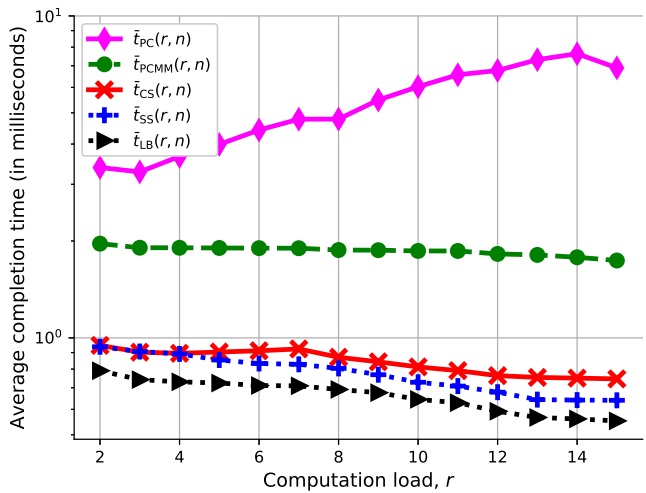}
\caption{Average completion time of different schemes with respect to computation load, $r \ge 2$.}
\label{AmazonFig_time_vs_r}
\end{figure}

Next, we present the results of experiments carried out on Amazon EC2 cluster. We compare the average completion time of different schemes with respect to the computation load $r$, $r \ge 2$, in Fig. \ref{AmazonFig_time_vs_r}, where $n = 15$, $d = 400$, and $N = 900$. As it can be seen, CS and SS outperform PC and PCMM significantly; while PCMM improves upon PC. This result shows that standard coded computation framework cannot fully exploit the computing capabilities in the network, and splitting the computational tasks assigned to each worker and receiving partial computations performed by each worker can reduce the average completion time significantly. We also observe that the average completion time of PC increases with $r$. This is because the delays at different workers are not significantly different; and thus, increasing the computation load to reduce the number of received computations from different workers can increase the total delay. This is another limitation of the coded computation framework, as it requires careful tuning of the parameters based on the statistics of the delays in the system. We observe that the gap between the average completion time of SS and the lower bound is relatively small for the entire range, and reduces with $r$, and SS outperforms CS with the improvement slightly increasing with $r$. The average completion time of RA, which requires $r = n$, is $\bar{t}_{{\rm{RA}}} (n,n) = 0.895$ millisecond, while SS achieves $\bar{t}_{{\rm{SS}}} (n,n) = 0.64$ millisecond, i.e., around $\% 28.5$ reduction. Thus, designing the TO matrix, rather than random computations, can provide significant improvement in computation speed. We observe that the average completion time of each scheme considered in Figures \ref{Fig_Exp_Dist_Scenario12} and \ref{AmazonFig_time_vs_r} follows a similar pattern. This verifies that the truncated Gaussian model can reasonably capture the statistical behaviour of the delays.            
%, in which the workers send their partial computations sequentially to the master,

\begin{figure}[t!]
\centering
\includegraphics[scale=0.6,trim={8pt 7pt 45pt 35pt},clip]{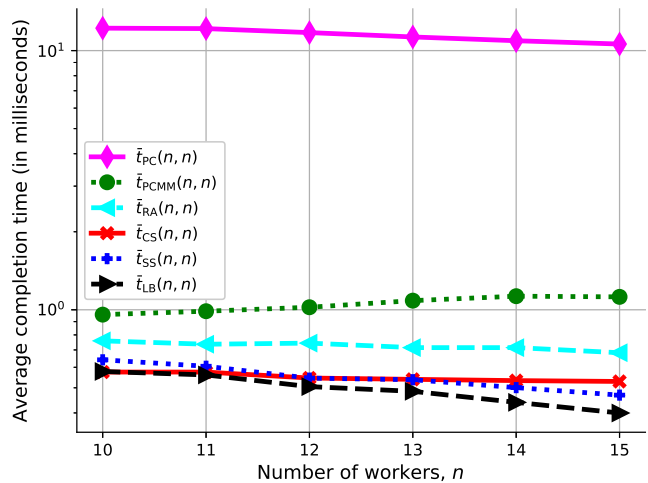}
\caption{Average completion time of different schemes with respect to the number of workers, $10 \le n \le 15$.}
\label{AmazonFig_time_vs_n}
\end{figure}

In Fig. \ref{AmazonFig_time_vs_n}, we compare the performances of different schemes with respect to the number of workers, $n$. We consider $d = 500$, $N = 1000$, and $r = n$. When $N / n$ is not an integer, we zero-pad the dataset. We observe that, except PCMM, the average completion time of different schemes reduce slightly with $n$ when $N$ is fixed. For PC, when $r = n$, the computation received from the fastest worker determines the completion time, and, with all other parameters fixed, the computation delay at each worker depends mostly on $N$. Thus, by introducing new workers when $N$ is fixed, the average completion time is expected to decrease. Whereas, with PCMM, although the computation time of each task is expected to decrease with $n$, the average completion time increases. This is due to the increase in the number of communications required by a factor of two as we have $\overline{t}_{\rm{PCMM}} (r,n) = \mathbb{E} \left[ t_{{\rm{PCMM}}, (2 n - 1)} \right]$. For uncoded computing schemes, RA, CS and SS, the average completion time decreases with $n$, as they allow a better utilization of the computing resources. As before, we observe that CS and SS improve the average completion time significantly compared to PC and PCMM. Also, based on the superiority of the CS and SS over RA, we conclude that the TO matrix design is essential in reducing the average delay of uncoded computing schemes. CS outperforms SS for small $n$ values, but SS takes over as $n$ increases. The relatively small gap between the average completion times of CS and SS and the lower bound illustrates their efficiency in scheduling the tasks despite the lack of any information on the speeds of the workers.

\begin{figure}[t!]
\centering
\includegraphics[scale=0.6,trim={0pt 6pt 45pt 35pt},clip]{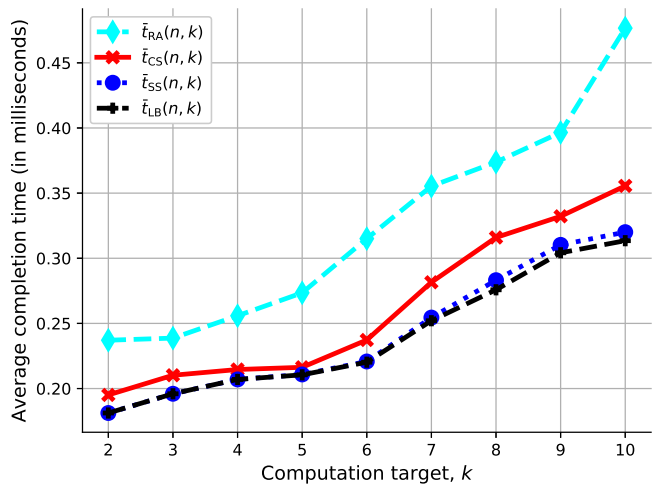}
\caption{Average completion time of different schemes with respect to the computation target, $2 \le k \le n$.}
\label{AmazonFig_time_vs_k}
\end{figure}

In Fig. \ref{AmazonFig_time_vs_k}, we compare the performance of different uncoded computation schemes and the lower bound with respect to the computation target, $k$. We set $n = 10$, $r = n$, $N = 1000$, and $d =800$, and consider $k \in [2:n]$. As expected, the average completion time increases with $k$. The gap between different schemes also increases with $k$, as the efficiency of scheduling tasks is more distinguishable for higher values of $k$. The average completion time of SS coincides with the lower bound for small and medium $k$ values $k \in [2:6]$, and the gap between the two is negligible even for higher $k$ values. Here we do not consider the coded computing schemes, PC and PCMM, as they are designed only for $k=n$.

%We compare the upper and lower bounds on the average completion time as a function of $k$ for $r=n$ in Scenario 7 and Scenario 8 shown in Fig. \ref{kScenario7} and Fig. \ref{kScenario8}, respectively. We note that, according to the considered framework, the PC and PCMM schemes are valid only for $k=n$; and hence their performances are represented with single pint at $k = n$. As expected, all the bounds are increasing with $k$. As illustrated in Fig. \ref{kScenarios78}, compared to Scenario 7, the CS and SS schemes perform better in Scenario 8, in which case the worker speeds are more diverse, and the gap between their performances and the lower bound is smaller. For lower $k$ values, the gap between the average completion times of CS and SS and the lower bound is relatively small, and it increases with $k$.     

\section{Conclusions}\label{SecConc}
We have studied distributed computation across inhomogeneous workers. The computation here may correspond to each iteration of a DGD algorithm applied on a large dataset, and it is considered to be completed when the master receives $k$ distinct computations. We assume that each worker has access to a limited portion of the dataset, defined as the computation load. In contrast to the growing literature on coded computation to mitigate straggling servers, here we have studied uncoded computations and sequential communication to the master in order to benefit from all the computations carried out by the workers, including the slower ones. Since the instantaneous computation speeds of the workers are not known in advance, allocation of the tasks to the workers and their scheduling become crucial in minimizing the average completion time. In particular, we have considered the assignment of data points to the workers with a predesigned computation order. Workers send the result of each computation to the master as soon as it is executed, and move on to compute the next task assigned to them. Assuming a general statistics for the computation and communication delays of different workers, we have obtained closed-form expressions for the average completion time of two particular computation allocation schemes, called CS and SS. The CS scheme dictates the same computation order at different workers, which is implemented by a cyclic shift operator. With SS, we introduce inverse computation orders at the workers. We have compared the performance of these proposed schemes with the existing ones in the literature, particularly the coded PC \cite{DCLiPolynomialCodedRegression}, PCMM \cite{DCEmrePCMM}, and uncoded RA \cite{DCLiRandomAssignment} schemes. The results of the experiments carried out on Amazon EC2 cluster show that the CS and SS schemes provide significant reduction in the average completion time over these schemes. The poor performance of the PC scheme can be explained by the fact that when the delays associated with different workers are not highly skewed, utilizing the partial computations by the slower workers becomes beneficial. The superiority of CS and SS compared to the RA scheme, which randomly schedules the tasks, illustrates the importance of task scheduling in speeding up the computations. 
%However, since the goal is to recover distinct computations at the server, having the same computation order at all the workers may delay some of the computations.

We also remark that, unlike the proposed schemes, the PC and PCMM schemes introduce additional encoding and decoding complexities at the master, which have not been considered in the evaluations here. Moreover, in the case of DSGD, having computed the partial gradient on separate data points may allow the workers to exploit more advanced methods to reduce their communication load, such as compression \cite{DCKonecnyFederated} or quantization \cite{DCOneBitQuan,Strom2015ScalableDD}, and can be beneficial in the case of communications over noisy channels \cite{DCMohammadDenizDSGDMACFederated}, which may not be applicable in the case of coded computations.

\appendices 
\section{Proof of Lemma \ref{LemmaHGGprime}}\label{AppProofInduction}
We prove the equation in \eqref{HGExpressionFori} by induction. For the ease of presentation, we define set $\mathcal{G}_i$ as $\mathcal{G}_i \subset [n]$ such that $\left| \mathcal{G}_i \right| = i$, and we denote $\mathcal{G}'_i = [n] \backslash \mathcal{G}_i$, $i \in [n]$. We first show that the equality in \eqref{HGExpressionFori} holds for any $\mathcal{G}'_{n-1}$ with $\left| \mathcal{G}'_{n-1} \right| = 1$ (note that the proof is trivial for $\mathcal{G}'_n$). According to \eqref{HGExpressionForiSimplification}, we have
\begin{align}\label{AppHGExpressionForiSimplificationGn_1}
H_{\mathcal{G}_{n-1},\mathcal{G}'_{n-1}} = H_{\mathcal{G}_{n-1}, \emptyset } - H_{ [n], \emptyset },
\end{align}
which is identical to the equality in \eqref{HGExpressionFori}. Then, assuming that for any $\mathcal{G}'_{i}$ with $\left| \mathcal{G}'_{i} \right| = n-i$, we have
\begin{align}\label{AppHGExpressionForiG_i}
H_{\mathcal{G}_i,\mathcal{G}'_i} = \sum\nolimits_{m=i}^{n} (-1)^{i+m} \sum\nolimits_{\hat{\mathcal{G}} \subset \mathcal{G}'_i: \left| \hat{\mathcal{G}} \right| = m-i} H_{\mathcal{G}_i \cup \hat{\mathcal{G}}, \emptyset},
\end{align}
we prove that 
\begin{align}\label{AppHGExpressionForiG_i_1}
H_{\mathcal{G}_{i-1},\mathcal{G}'_{i-1}} = & \sum\nolimits_{m=i}^{n} (-1)^{i+m-1} \cdot \nonumber\\
& \qquad \quad \sum\nolimits_{\hat{\mathcal{G}} \subset \mathcal{G}'_{i-1}: \left| \hat{\mathcal{G}} \right| = m-i+1} H_{\mathcal{G}_{i-1} \cup \hat{\mathcal{G}}, \emptyset}.
\end{align}
Consider a fixed set $\mathcal{G}_{i}$. For $g_i \in \mathcal{G}_i$, let $\mathcal{G}_{i-1} = \mathcal{G}_i \backslash \{ g_i \}$, which results in $\mathcal{G}'_{i-1} = \mathcal{G}'_i \cup \{ g_i \}$. From \eqref{HGExpressionForiSimplification}, we have   
\begin{align}\label{AppHGExpressionForiSimplificationGprime_i_1}
H_{\mathcal{G}_{i-1},\mathcal{G}'_{i-1}} &= H_{\mathcal{G}_{i-1},\mathcal{G}'_{i-1} \backslash \left\{ g_i \right\} } - H_{ \mathcal{G}_{i-1} \cup \{ g_i \},\mathcal{G}'_{i-1} \backslash \left\{ g_i \right\} } \nonumber\\
& = H_{\mathcal{G}_{i} \backslash \{ g_i \},\mathcal{G}'_{i} } - H_{ \mathcal{G}_{i}, \mathcal{G}'_{i} }.
\end{align}
According to \eqref{AppHGExpressionForiG_i}, it follows that
\begin{align}\label{AppProofPart1}
& H_{\mathcal{G}_{i} \backslash \{ g_i \},\mathcal{G}'_{i} } \nonumber\\
& = \sum\nolimits_{m=i-1}^{n-1} (-1)^{i+m-1} \sum\nolimits_{\hat{\mathcal{G}} \subset \mathcal{G}'_i: \left| \hat{\mathcal{G}} \right| = m-i+1} H_{\mathcal{G}_{i-1} \cup \hat{\mathcal{G}}, \emptyset} \nonumber\\
& = \sum\nolimits_{m=i-1}^{n} (-1)^{i+m-1} \sum\nolimits_{\hat{\mathcal{G}} \subset \mathcal{G}'_i: \left| \hat{\mathcal{G}} \right| = m-i+1} H_{\mathcal{G}_{i-1} \cup \hat{\mathcal{G}}, \emptyset},
\end{align}
and 
\begin{align}\label{AppProofPart2}
& H_{\mathcal{G}_{i},\mathcal{G}'_{i} } = \sum\nolimits_{m=i}^{n} (-1)^{i+m} \sum\nolimits_{\hat{\mathcal{G}} \subset \mathcal{G}'_i: \left| \hat{\mathcal{G}} \right| = m-i} H_{\mathcal{G}_{i} \cup \hat{\mathcal{G}}, \emptyset} \nonumber\\
& = \sum\nolimits_{m=i-1}^{n} (-1)^{i+m} \sum\nolimits_{\hat{\mathcal{G}} \subset \mathcal{G}'_i: \left| \hat{\mathcal{G}} \right| = m-i} H_{\mathcal{G}_{i} \cup \hat{\mathcal{G}}, \emptyset} \nonumber\\
& = \sum\nolimits_{m=i-1}^{n} (-1)^{i+m} \sum\nolimits_{\hat{\mathcal{G}} \subset \mathcal{G}'_i: \left| \hat{\mathcal{G}} \right| = m-i} H_{\mathcal{G}_{i-1} \cup \{ g_i \} \cup \hat{\mathcal{G}}, \emptyset}.
\end{align}
By plugging \eqref{AppProofPart1} and \eqref{AppProofPart2} in \eqref{AppHGExpressionForiSimplificationGprime_i_1}, we have 
\begin{align}\label{AppProofPart12}
& H_{\mathcal{G}_{i-1},\mathcal{G}'_{i-1}} \nonumber\\
& = \sum\nolimits_{m=i-1}^{n} (-1)^{i+m-1} \bigg(  \sum\nolimits_{\hat{\mathcal{G}} \subset \mathcal{G}'_i: \left| \hat{\mathcal{G}} \right| = m-i+1} H_{\mathcal{G}_{i-1} \cup \hat{\mathcal{G}}, \emptyset} \bigg. \nonumber\\
& \qquad \qquad \qquad \quad \bigg. + \sum\nolimits_{\hat{\mathcal{G}} \subset \mathcal{G}'_i: \left| \hat{\mathcal{G}} \right| = m-i} H_{\mathcal{G}_{i-1} \cup \{ g_i \} \cup \hat{\mathcal{G}}, \emptyset} \bigg) \nonumber\\
& = \sum\nolimits_{m=i-1}^{n} (-1)^{i+m-1} \sum\nolimits_{\hat{\mathcal{G}} \subset \mathcal{G}'_i \cup \{ g_i \}: \left| \hat{\mathcal{G}} \right| = m-i+1} H_{\mathcal{G}_{i-1} \cup \hat{\mathcal{G}}, \emptyset} \nonumber\\
& = \sum\nolimits_{m=i-1}^{n} (-1)^{i+m-1} \sum\nolimits_{\hat{\mathcal{G}} \subset \mathcal{G}'_{i-1}: \left| \hat{\mathcal{G}} \right| = m-i+1} H_{\mathcal{G}_{i-1} \cup \hat{\mathcal{G}}, \emptyset},
\end{align}
which provides the proof of the equality in \eqref{AppHGExpressionForiG_i_1}. This completes the proof of \eqref{HGExpressionFori}.

%\vspace{-.1cm}
\bibliographystyle{IEEEtran}
\bibliography{Report}

\begin{thebibliography}{10}
\providecommand{\url}[1]{#1}
\csname url@samestyle\endcsname
\providecommand{\newblock}{\relax}
\providecommand{\bibinfo}[2]{#2}
\providecommand{\BIBentrySTDinterwordspacing}{\spaceskip=0pt\relax}
\providecommand{\BIBentryALTinterwordstretchfactor}{4}
\providecommand{\BIBentryALTinterwordspacing}{\spaceskip=\fontdimen2\font plus
\BIBentryALTinterwordstretchfactor\fontdimen3\font minus
  \fontdimen4\font\relax}
\providecommand{\BIBforeignlanguage}[2]{{%
\expandafter\ifx\csname l@#1\endcsname\relax
\typeout{** WARNING: IEEEtran.bst: No hyphenation pattern has been}%
\typeout{** loaded for the language `#1'. Using the pattern for}%
\typeout{** the default language instead.}%
\else
\language=\csname l@#1\endcsname
\fi
#2}}
\providecommand{\BIBdecl}{\relax}
\BIBdecl

\bibitem{DCBoyd}
S.~Boyd, N.~Parikh, E.~Chu, B.~Peleato, and J.~Eckstein, ``Distributed
  optimization and statistical learning via the alternating direction method of
  multipliers,'' \emph{Foundations and Trends in Machine Learning}, vol.~3,
  no.~1, pp. 1--122, 2012.

\bibitem{DCChoi}
J.~Choi, D.~W. Walker, and J.~J. Dongarra, ``Pumma: {Parallel} universal matrix
  multiplication algorithms on distributed memory concurrent computers,''
  \emph{Concurrency: Practice and Experience}, vol.~6, no.~7, pp. 543--570,
  1994.

\bibitem{DCDisMachLearn}
K.~Lee, M.~Lam, R.~Pedarsani, D.~Papailiopoulos, and K.~Ramchandran, ``Speeding
  up distributed machine learning using codes,'' \emph{{IEEE} Trans. Inform.
  Theory}, vol.~64, no.~3, pp. 1514--1529, Mar. 2018.

\bibitem{DCTandonAlexGD}
R.~Tandon, Q.~Lei, A.~G. Dimakis, and N.~Karampatziakis, ``Gradient coding:
  {Avoiding} stragglers in distributed learning,'' in \emph{Proc. of ICML},
  Sydney, Australia, Aug. 2017, pp. 3368--3376.

\bibitem{DCReedSol}
W.~Halbawi, N.~A. Ruhi, F.~Salehi, and B.~Hassibi, ``Improving distributed
  gradient descent using {Reed-Solomon} codes,'' \emph{{arXiv:1706.05436v1
  [cs.IT]}}, Jun. 2017.

\bibitem{DCRaceDutta}
S.~Dutta, G.~Joshi, S.~Ghosh, P.~Dube, and P.~Nagpurkar, ``Slow and stale
  gradients can win the race: {Error}-runtime trade-offs in distributed
  {SGD},'' \emph{{arXiv:1803.01113v3 [cs.IT]}}, May 2018.

\bibitem{DCMomentLDPC}
R.~K. Maity, A.~S. Rawat, and A.~Mazumdar, ``Robust gradient descent via moment
  encoding with {LDPC} codes,'' \emph{{arXiv:1805.08327v1 [cs.IT]}}.

\bibitem{DCAbbeCC}
M.~Ye and E.~Abbe, ``Communication-computation efficient gradient coding,''
  \emph{{arXiv:1802.03475v1 [cs.IT]}}, Feb. 2018.

\bibitem{DCYuMaddahMatMult}
Q.~Yu, M.~A. Maddah-Ali, and A.~S. Avestimehr, ``Straggler mitigation in
  distributed matrix multiplication: {Fundamental} limits and optimal coding,''
  \emph{{arXiv:1801.07487v2 [cs.IT]}}, May 2018.

\bibitem{DCDuttaFahimHadMatMult}
S.~Dutta, M.~Fahim, F.~Haddadpour, H.~Jeong, V.~Cadambe, and P.~Grover, ``On
  the optimal recovery threshold of coded matrix multiplication,''
  \emph{{arXiv:1801.10292v2 [cs.IT]}}, May 2018.

\bibitem{QianPolynHighDimMatMat}
Q.~Yu, M.~A. {Maddah-Ali}, and A.~S. Avestimehr, ``Polynomial codes: an optimal
  design for high-dimensional coded matrix multiplication,''
  \emph{{arXiv:1705.10464 [cs.IT]}}, Jan. 2018.

\bibitem{DCYuResiliencySecurityPrivacy}
Q.~Yu, N.~Raviv, J.~So, and A.~S. Avestimehr, ``Lagrange coded computing:
  {Optimal} design for resiliency, security and privacy,''
  \emph{{arXiv:1806.00939v2 [cs.IT]}}, Jun. 2018.

\bibitem{DCLiPolynomialCodedRegression}
S.~Li, S.~M. {Mousavi Kalan}, Q.~Yu, M.~Soltanolkotabi, and A.~S. Avestimehr,
  ``Polynomially coded regression: {Optimal} straggler mitigation via data
  encoding,'' \emph{{arXiv:1805.09934v1 [cs.IT]}}, May 2018.

\bibitem{DCHierarchical}
N.~Ferdinand and S.~C. Draper, ``Hierarchical coded computation,'' in
  \emph{Proc. {IEEE} Int'l Symp. on Inform. Theory (ISIT)}, Vail, CO, USA, Jun.
  2018, pp. 1620--1624.

\bibitem{DCKianiStragglers}
S.~Kiani, N.~Ferdinand, and S.~C. Draper, ``Exploitation of stragglers in coded
  computation,'' \emph{{arXiv:1806.10253v1 [cs.IT]}}, Jun. 2018.

\bibitem{DCRateless}
A.~Mallick, M.~Chaudhari, and G.~Joshi, ``Rateless codes for near-perfect load
  balancing in distributed matrix-vector multiplication,''
  \emph{{arXiv:1804.10331v2 [cs.DC]}}, Mar. 2018.

\bibitem{DCEmrePCMM}
E.~Ozfatura, D.~G\"und\"uz, and S.~Ulukus, ``Speeding up distributed gradient
  descent by utilizing non-persistent stragglers,'' \emph{{arXiv:1808.02240v2
  [cs.IT]}}, Aug. 2018.

\bibitem{DCLiRandomAssignment}
S.~Li, S.~M. {Mousavi Kalan}, A.~S. Avestimehr, and M.~Soltanolkotabi,
  ``Near-optimal straggler mitigation for distributed gradient methods,''
  \emph{{arXiv:1710.09990v1 [cs.IT]}}, Oct. 2017.

\bibitem{DCMohammadDenizDSGDMACFederated}
{M. Mohammadi Amiri} and D.~G\"und\"uz, ``Machine learning at the wireless
  edge: {Distributed} stochastic gradient descent over-the-air,''
  \emph{{arXiv:1901.00844 [cs.DC]}}, Feb. 2019.

\bibitem{DCKonecnyFederated}
J.~Konecny, H.~B. McMahan, F.~X. Yu, P.~Richtarik, A.~T. Suresh, and D.~Bacon,
  ``Federated learning: {Strategies} for improving communication efficiency,''
  \emph{{arXiv:1610.05492v2 [cs.LG]}}, Oct. 2017.

\bibitem{DCOneBitQuan}
F.~Seide1, H.~Fu, J.~Droppo, G.~Li, and D.~Yu, ``1-bit stochastic gradient
  descent and its application to data-parallel distributed training of speech
  {DNN}s,'' in \emph{INTERSPEECH}, Singapore, Sep. 2014, pp. 1058--1062.

\bibitem{Strom2015ScalableDD}
N.~Strom, ``Scalable distributed {DNN} training using commodity {GPU} cloud
  computing,'' in \emph{INTERSPEECH}, 2015, pp. 1488--1492.

\bibitem{AttiaRaviUncodedDC}
M.~A. Attia and R.~Tandon, ``Combating computational heterogeneity in
  large-scale distributed computing via work exchange,''
  \emph{{arXiv:1711.08452 [cs.DC]}}, Nov. 2017.

\bibitem{JobSchedulingGraham}
R.~L. Graham, ``Bounds on multiprocessing anomalies and packing algorithms,''
  in \emph{Proc. SJCC}, 1972, pp. 205--218.

\bibitem{UllmanNPCpmpleteScheduling}
J.~D. Ullman, ``{NP}-complete scheduling problems,'' \emph{J. Comput. Syst.
  Sci.}, vol.~10, no.~3, pp. 384--393, Jun. 1975.

\bibitem{XueSchedCloudCompACOLB}
S.~Xue, M.~Li, X.~Xu, and J.~Chen, ``An {ACO-LB} algorithm for task scheduling
  in the cloud environment,'' \emph{J. Software}, vol.~9, no.~2, pp. 466--473,
  Feb. 2014.

\bibitem{ZhanPCOCpmpleteScheduling}
S.~Zhan and H.~Huo, ``Improved {PSO}-based task scheduling algorithm in cloud
  computing,'' \emph{J. Inf. Comput. Sci.}, vol.~9, no.~13, pp. 3821--3829,
  Nov. 2012.

\bibitem{ReddySchedCloudCompMultiObj}
G.~Reddy, N.~Reddy, and S.~Phanikumar, ``Multi objective task scheduling using
  modified ant colony optimization in cloud computing,'' \emph{Int'l J. Intell.
  Eng. Sys.}, vol.~11, no.~3, pp. 242--250, 2018.

\bibitem{ChoudhariPriorFogComp}
T.~Choudhari, M.~Moh, and T.-S. Moh, ``Prioritized task scheduling in fog
  computing,'' in \emph{Proc. ACMSE}, Richmond, Kentucky, Mar. 2018, pp.
  401--405.

\bibitem{YinFogComputeSmartManufacture}
L.~Yin, J.~Luo, and H.~Luo, ``Tasks scheduling and resource allocation in fog
  computing based on containers for smart manufacturing,'' \emph{{IEEE} Trans.
  ON Indus. Inform.}, vol.~14, no.~10, pp. 4712--4721, Oct. 2018.

\bibitem{PedarsaniSchedulingMultServers}
R.~Pedarsani, J.~Walrand, and Y.~Zhong, ``Scheduling tasks with precedence
  constraints on multiple servers,'' in \emph{Proc. Allerton}, Monticello, IL,
  2014, pp. 1196--1203.

\bibitem{PedarsaniAverstimehrSchedulingComAware}
C.-S. Yang, R.~Pedarsani, and A.~S. Avestimehr, ``Communication-aware
  scheduling of serial tasks for dispersed computing,'' \emph{{arXiv:1804.06468
  [cs.DC]}}, Apr. 2018.

\bibitem{DCAjiSparse}
A.~F. Aji and K.~Heafield, ``Sparse communication for distributed gradient
  descent,'' \emph{{arXiv:1704.05021v2 [cs.CL]}}, Jul. 2017.

\bibitem{DCSattlerSparseBinary}
F.~Sattler, S.~Wiedemann, K.-R. M\"uller, and W.~Samek, ``Sparse binary
  compression: {Towards} distributed deep learning with minimal
  communication,'' \emph{{arXiv:1805.08768v1 [cs.LG]}}, May 2018.

\bibitem{DCChenAdaComp}
C.-Y. Chen, J.~Choi, D.~Brand, A.~Agrawal, W.~Zhang, and K.~Gopalakrishnan,
  ``{AdaComp} : {Adaptive} residual gradient compression for data-parallel
  distributed training,'' \emph{{arXiv:1712.02679v1 [cs.LG]}}, Dec. 2017.

\bibitem{DCLinHanDeepGradComp}
Y.~Lin, S.~Han, H.~Mao, Y.~Wang, and W.~J. Dally, ``Deep gradient compression:
  {Reducing} the communication bandwidth for distributed training,''
  \emph{{arXiv:1712.01887v2 [cs.CV]}}, Feb. 2018.

\bibitem{DCTaoLieSGD}
Z.~Tao and Q.~Li, ``{eSGD}: {Communication} efficient distributed deep learning
  on the edge,'' in \emph{Workshop on Hot Topics in Edge Computing (HotEdge)},
  Boston, MA, USA, Jul. 2018.

\bibitem{DCWangATOMO}
H.~Wang, S.~Sievert, S.~Liu, Z.~Charles, D.~Papailiopoulos, and S.~Wright,
  ``{ATOMO}: {Communication}-efficient learning via atomic sparsification,''
  \emph{{arXiv:1806.04090v2 [stat.ML]}}, Jun. 2018.

\end{thebibliography}

\end{document}